\documentclass[]{aa}
\usepackage{graphics}
\usepackage{latexsym}
\input epsf
\begin{document}

\sloppy

\thesaurus{(08.08.1;08.19.1;10.19.2;10.19.3)}
   
\title{The Galactic disk: study of four low latitude Galactic Fields}
%\subtitle{XXX}
\thanks{Based on observations collected at the European
Southern Observatory, La Silla, Chile}

\author{A. Vallenari \inst{1}, G. Bertelli \inst{2}, L. Schmidtobreick \inst{1}       }

\offprints{A. Vallenari}

\institute{Padova Astronomical Observatory, vicolo Osservatorio 
          5, I-35122, Padova,
        Italy
\and
Department of Astronomy, Padova University,
        vicolo dell'Osservatorio 5, I-35122, Padova, Italy\\
e-mail: {\tt 
vallenari\char64pd.astro.it, bertelli\char64pd.astro.it, 
linda\char64pd.astro.it
}}

\date{Received May 2000; accepted July 2000}

\maketitle

\markboth{A. Vallenari et al.}{The Galactic Disk}

\begin{abstract}
We present  deep V and I  photometry of stellar field 
in four previously unstudied low latitude regions of the Galactic disk.
All observed fields are located at the western side of the Galactic Center 
in the direction
of the Coalsack-Carina region.
They are chosen on the large scale surface photometry
of the Milky Way (Hoffmann et al., 1998 and Kimeswenger et al., 1993) and
corresponding term maps 
 (Schlosser et al., 1995) as being affected by low interstellar absorption
and having integrated colours typical of a very young population.
Two of them are suspected to cross the inner spiral arm. 
More than $10^5$ stars are detected in total, 
down to a magnitude of V $\sim 23.5$. 
The observational  colour-magnitude diagrams (CMDs) and luminosity functions
are 
analyzed using  a revised version of the
 Padova software  described in Bertelli et al (1995).
The interstellar extinction along the line of sight is derived and
 found to be in reasonable agreement  with \cite{Men+98} maps.
 Due to the low galactic latitude of the studied fields,
the scale length and mainly the scale height of the thick disk are not strongly
constrained by the observations. However  
a  thin disk scale height of about 250$\pm 60$ pc  seems to be favored.
The data are very sensitive to the star formation rate of the thin disk.
A decreasing star formation rate  is 
necessary to reproduce the distribution of the stars in the  colour-magnitude diagrams as well as the luminosity functions.
A constant or a strongly increasing star formation rate  
as  derived using Hipparcos data for the solar
neighborhood (Bertelli et al 1999)
are marginally in agreement with
the luminosity functions but they are at odds with the CMDs.
 The analysis of these data suggest that the  solar
neighborhood  star formation rate  cannot be considered as representative
for the whole thin disk. To properly reproduce the luminosity functions
a thick disk component  having  a  local density  of about 2--4\%  must be included. From the star-counts the local neighborhood mass density in stars more massive than 0.1 M$_\odot$ 
 is found to be 0.036--0.02 M$_\odot$ pc $^{-3}$.
Finally, the location of inner spiral arm is discussed. We find
evidence  
 of a population younger than 10$^8$ yr distributed in a  spiral arm  at a distance 
of 1.3$\pm 0.2$ and 1.5 $\pm 0.2$\,Kpc  in the directions
  l$\sim$\,292 and l$\sim$\,305
respectively. This result is consistent with the
spiral arm pattern defined on the basis of pulsars and
young associations
(\cite{TayCor93}; \cite{Hum76}).
Due to the small field of view two of the studied fields
do not set strong constrains on the scale height and lenght of the disk.
A  larger field of view, see e.g.
the WFI at the 2.2m ESO telescope having 30$^\prime \times$ 30$^\prime$
would allow us to have good statistics down to faint magnitudes.

\end{abstract}

\section{Introduction}
During the past years many efforts have been undertaken to gain information
about the structure of the Milky Way. However, several unresolved questions
 still remain. 
A program has been recently started to study the structure of the Galaxy
by means of deep photometry of stellar populations.  
Several fields in the direction of  the Galactic center 
has been been analyzed with the aim of 
 studying the Galactic structure; deriving 
 the stellar extinction along the line of sight; 
obtaining information about the age and the metallicity of the stars 
in various Galactic components (halo, bulge and disc);
deriving the past history of star formation and 
chemical enrichment.
 
In brief,  Ng et al. (1995) studied the color 
magnitude diagram (CMD) and luminosity functions (LF) of the
 field \#3 of the Palomar
Groningen Survey (PG3) located at the periphery of the Galactic Bulge 
(l=0, b=-10) and set up the basic tool to deconvolve the contribution from 
many stellar population to highly composite CMDs.  Bertelli et al. (1995)  
  analyzed three regions of the Bulge located  near 
the clusters NGC6603 (l=12.9 b=-2.8 ), Lynga 7 (l=328.8 b=-2.8) and Terzan 1 
(l=357.6 b=1.0). First they  show that
 the  extinction law greatly varies with the direction towards the center.
Second, they trace the position of a molecular ring between 3.5 and 4\,kpc in the 
direction of NGC6603, of a stellar ring between 3.0 and 4.0\,kpc, of the 
Sagittarius arm in the direction of NGC6603 (5.0 to 7.0\,kpc) and Lynga 7 
(4.2 to 7.0\,kpc). Third,  they recognize the
 presence near the Galactic Center of an old and metal 
rich population (see also the detailed study of the CMD of Terzan 1 by 
Bertelli et al. 1996 in which evidences for the existence of old, high 
metallicity stars in the so-called hot horizontal branch stage are found), 
and finally  give a hint that the Galactic Bar points its nearest side 
toward positive galactic longitude. 

This paper is devoted to the study of four  low latitude
fields in the Galactic disk
in the direction
of the Coalsack-Carina region. 
Studying the CMDs and luminosity functions
it is possible to determine the age and star formation history
of the thin disk components, the extinction  along the line of sight 
derive hints about the  
scale height and length.

In section \ref{sec_data} and in section \ref{sec_reduction}
the observations and 
 the data 
reduction respectively are described. In section
\ref{sec_cmd}
  the   CMDs are presented. 
 In section \ref{sec_model} the model and
the input parameters are given. In section \ref{sec_results} the results
are discussed. The presence of a spiral arm at l$\sim$\,292 and l$\sim$\,305
is analyzed in section  \ref{sec_spiral}. Finally, the conclusions are
drawn in section \ref{sec_conclusion}.

\begin{table*}[tb]
\caption{\label{coordinates} General information on the fields: central 
coordinates, number of exposures and integrated exposure time for both filters, 
dimension of the selected part of the field (see text for details), and total number of stars therein.}
\begin{tabular}{ c c c c c c c c c c c }
\hline
 \noalign{\smallskip}
& $\alpha_{2000}$ & $\delta_{2000}$ & $l$ & $b$ &
 $N_E(\mbox{V}) $ & $t_E(\mbox{V})/s$ &
 $N_E(\mbox{I}) $ & $t_E(\mbox{I})/s$ & 
 $size$ & $N_{stars}$  \\
 \noalign{\smallskip}
\hline
 \noalign{\smallskip}
Field 1 (F1) & $\,\,\,8^h40^m54^s$ &  $-46^\circ 56^\prime 40^{\prime\prime}$ &
 265.545 & $-3.089$ &
 6 & 1440 & 6 & 1080 & $6^{\prime}.4 \times 9^{\prime}.6$ & 2787\\
Field 2 (F2)& $\,\,\,9^h38^m54^s$ & $-53^\circ 13^\prime 00^{\prime\prime}$ &
276.448 & $-0.546$ & 
 10 & 1800 & 10 & 1200 & $12{\prime}.0 \times 12^{\prime}.0$ & 6580\\
Field 3 (F3)& $11^h27^m33^s$ & $-59^\circ 31^\prime 09^{\prime\prime}$ &
 292.468 & $\,\,\,\,1.638$ & 
 6 & 1080 & 4 & 720 & $9^{\prime}.6 \times 9^{\prime}.6$ & 12153\\
Field 4 (F4)& $13^h13^m27^s$ & $-67^\circ 39^\prime 29^{\prime\prime}$ &
 305.029 & $-4.876$ & 
 5 & 1500 & 5 & 1200 & $9^{\prime}.6 \times 9^{\prime}.6$ & 9380\\
\noalign{\smallskip}
\hline 
\end{tabular}
\end{table*}

\begin{figure}[t]
%\vspace{-8mm}
%\rotatebox{270}{\resizebox{4.5cm}{!}{\includegraphics{find.ps}}}
%\resizebox{8.7cm}{!}{\includegraphics{find.ps}}
\resizebox{8.7cm}{!}{\includegraphics{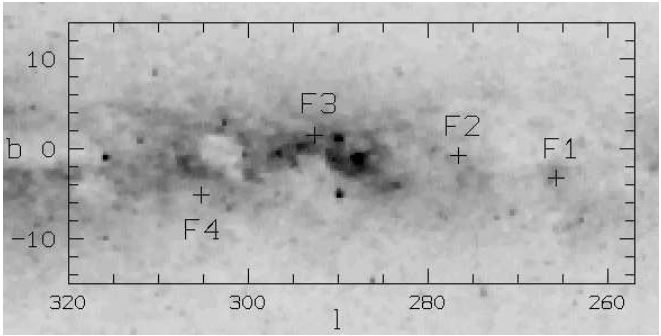}}
%\special{psfile=find.ps
%         hoffset=-30 voffset=00 hscale=37 vscale=37 angle=-90}
%\vspace{53mm}
\caption{\label{find} The location of the observed fields are indicated
in the Galactic map (V passband) of the Coalsack-Carina region}
\end{figure}

\section{Data}
\label{sec_data}
For this work, four fields have been observed in Bessel V and Gunn I with
the 1.54\,m Danish telescope at the European Southern Observatory (ESO),
La Silla (Chile) on February $\rm 10^{th}-13^{th}$, 1997
(camera C1W11/4 with 2k$\times$2k pixels and a nominal
field of view of $13^\prime\times13^\prime$).
The  coordinates of the fields,
the number of averaged exposures,
and the total exposure time per filter are given in Table \ref{coordinates}.

The fields have been chosen on the large scale surface photometry maps
of the Milky Way (Hoffmann et al., 1998 and Kimeswenger et al., 1993) and
corresponding term maps which display all points of the Milky Way
having the same colour (Schlosser et al., 1995). These maps show large
scale structures  and are therefore suitable for the
selection of interesting areas. The integrated colours of the fields
we select show they are affected by low interstellar absorption.
Two of the  fields point in the direction of the Sgr--Car arm and 
their integrated UBVR colors  are consistent with the presence 
of a 
significantly fraction young population.  
In Fig. \ref{find}, the positions of the 
observed fields are plotted over the relevant part of the Galactic V map.

\section{Reductions}
\label{sec_reduction}
All of the reductions have been carried out with the image processing system
MIDAS. For the stellar photometry, the DAOPHOT and ALLSTAR
code have been used.
The raw data has been corrected for bias and flat-fields. 
Due to technical
problems with the filter wheel resulting in vignetting, 
the flat-fields are not only very inhomogeneous
but do also vary in time. 
Therefore 
 some of the data couldn't 
be used at all. The dimensions of the unvignetted section of each field are given in
table \ref{coordinates}.

For the calibration, at least 10 standard stars
have been measured at the beginning, the middle and the end of every night
at different air-masses. They have been selected from Landolt (1992).
From the measured standard stars we derive the following relation
between instrumental v, (v-i) and calibrated V , (V-I) for 1 sec of exposure
time:

\begin{equation}
\label{calibration1}
\begin{array}{r@{\;\; = \;\;}l}
 V-I & 1.034\, (v-i) + 0.9824 \\
 V & v + 0.007\, (V-I)  - 1.26\\
%% I & i - 0.026\, (V-I) - 2.21 
\end{array}
\end{equation}

\begin{figure}[bt]
%\vspace{-8mm}
%\special{psfile=figcomp2b.ps
%         hoffset=-30 voffset=00 hscale=37 vscale=37 angle=-90}
%\vspace{53mm}
%{\resizebox{9.0cm}{!}{\includegraphics{figcomp2.ps}}}
{\resizebox{9.0cm}{!}{\includegraphics{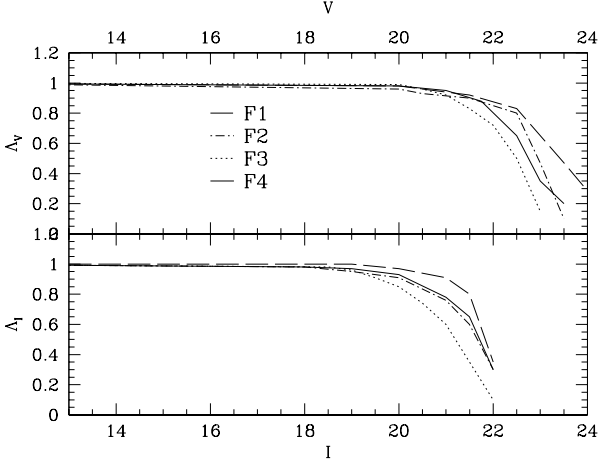}}}
\caption{\label{comple} Completeness factors $\Lambda_V$ and $\Lambda_I$
as  functions of the magnitudes for all the fields.}
\end{figure}

The  completeness corrections $\Lambda_V$ and $\Lambda_I$
defined as the ratios of recovered to injected stars in V and I frames
respectively  are derived by means of the usual
artificial stars experiments.
They are plotted for each field
and filter in Fig. \ref{comple}. The photometric errors
are  0.01,0.05,0.12,0.20 for V=10,21,22,23 
respectively and 0.01,0.02,0.05,0.2 for I=10,18,20,21.5 mag.

\section{Colour-Magnitude Diagrams (CMDs)}

\label{sec_cmd}
\begin{figure*}[t]
\parbox{8.7cm}{
%\resizebox{8.7cm}{!}{\includegraphics{f1n1.ps}}\\
\resizebox{8.7cm}{!}{\includegraphics{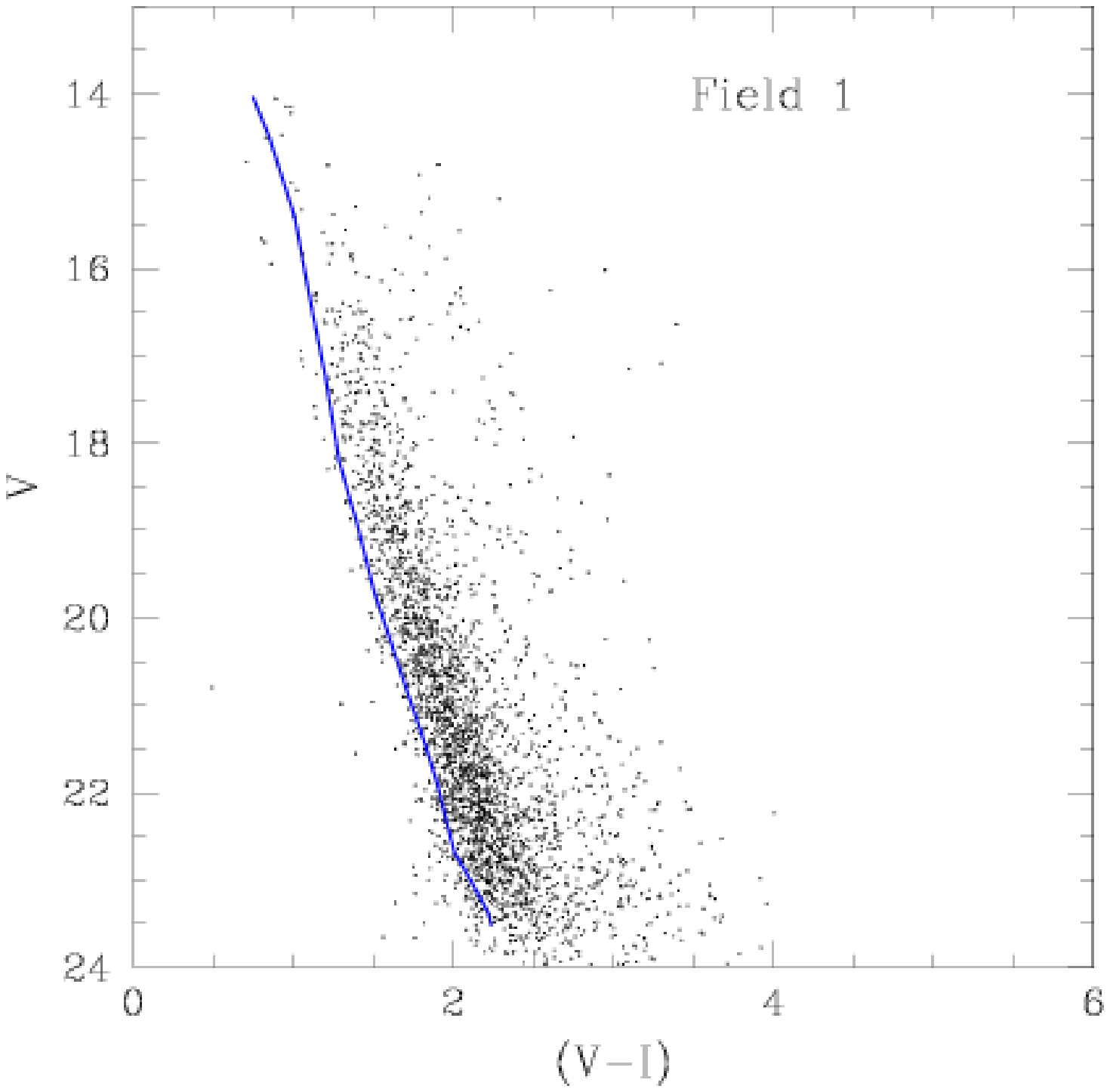}}\\
a)}
\hspace{0.5cm}
\parbox{8.7cm}{
%\resizebox{8.7cm}{!}{\includegraphics{f2n1.ps}}\\
\resizebox{8.7cm}{!}{\includegraphics{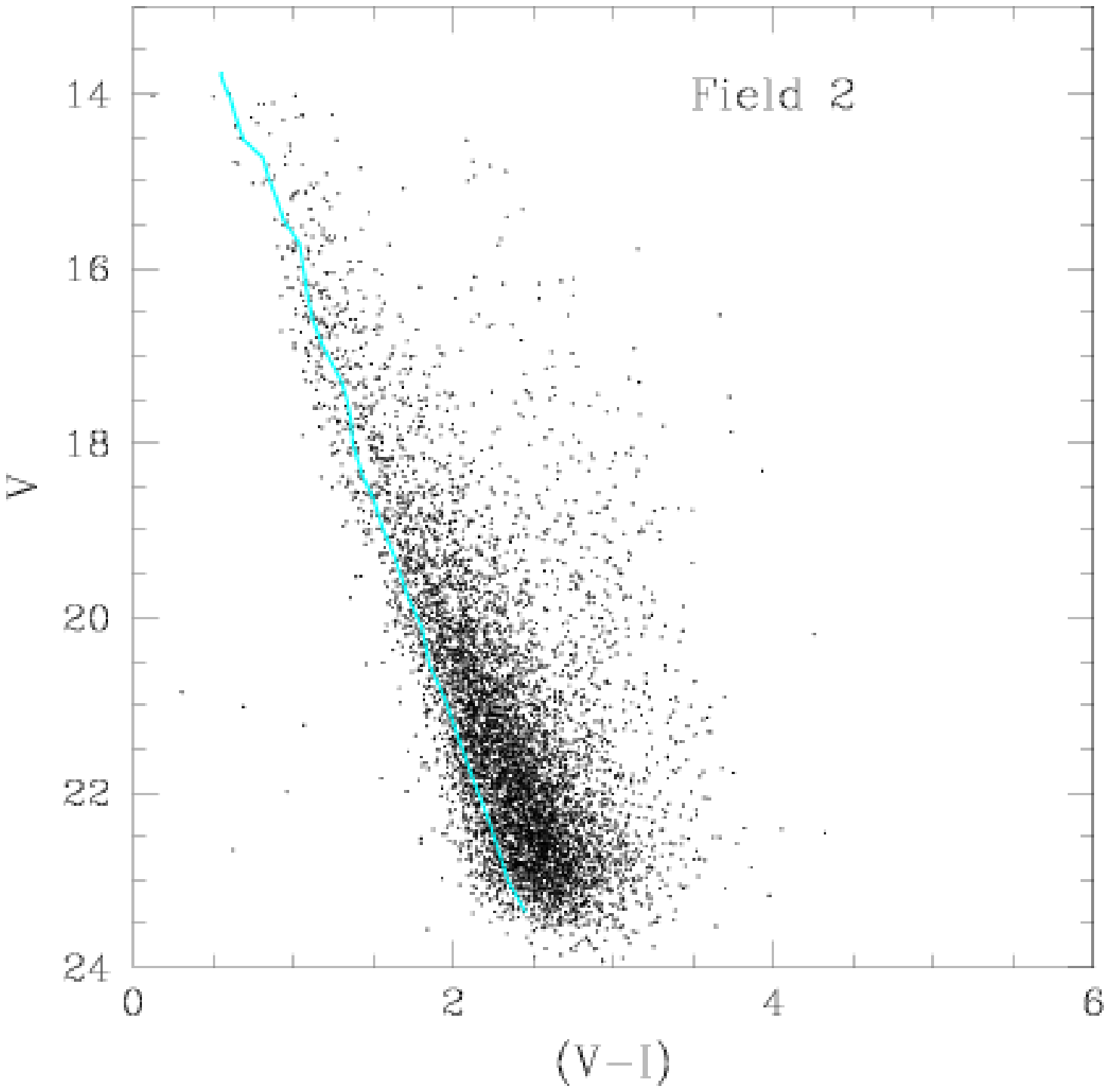}}\\
b)}
\hspace{0.5cm}
\parbox{8.7cm}{
%\resizebox{8.7cm}{!}{\includegraphics{f3n1.ps}}\\
\resizebox{8.7cm}{!}{\includegraphics{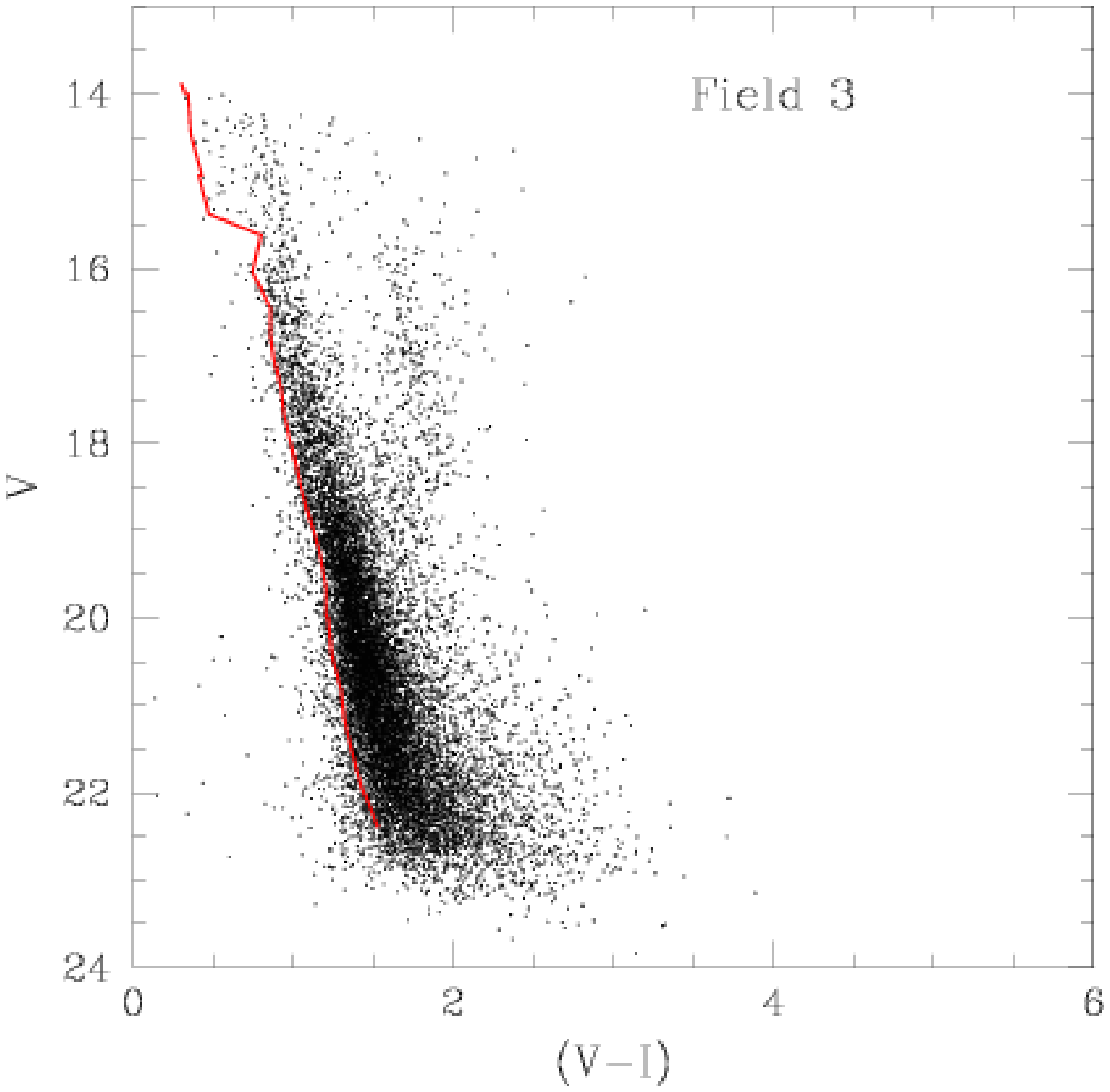}}\\
c)}
\hspace{0.5cm}
\parbox{8.7cm}{
%\resizebox{8.7cm}{!}{\includegraphics{f4n1.ps}}\\
\resizebox{8.7cm}{!}{\includegraphics{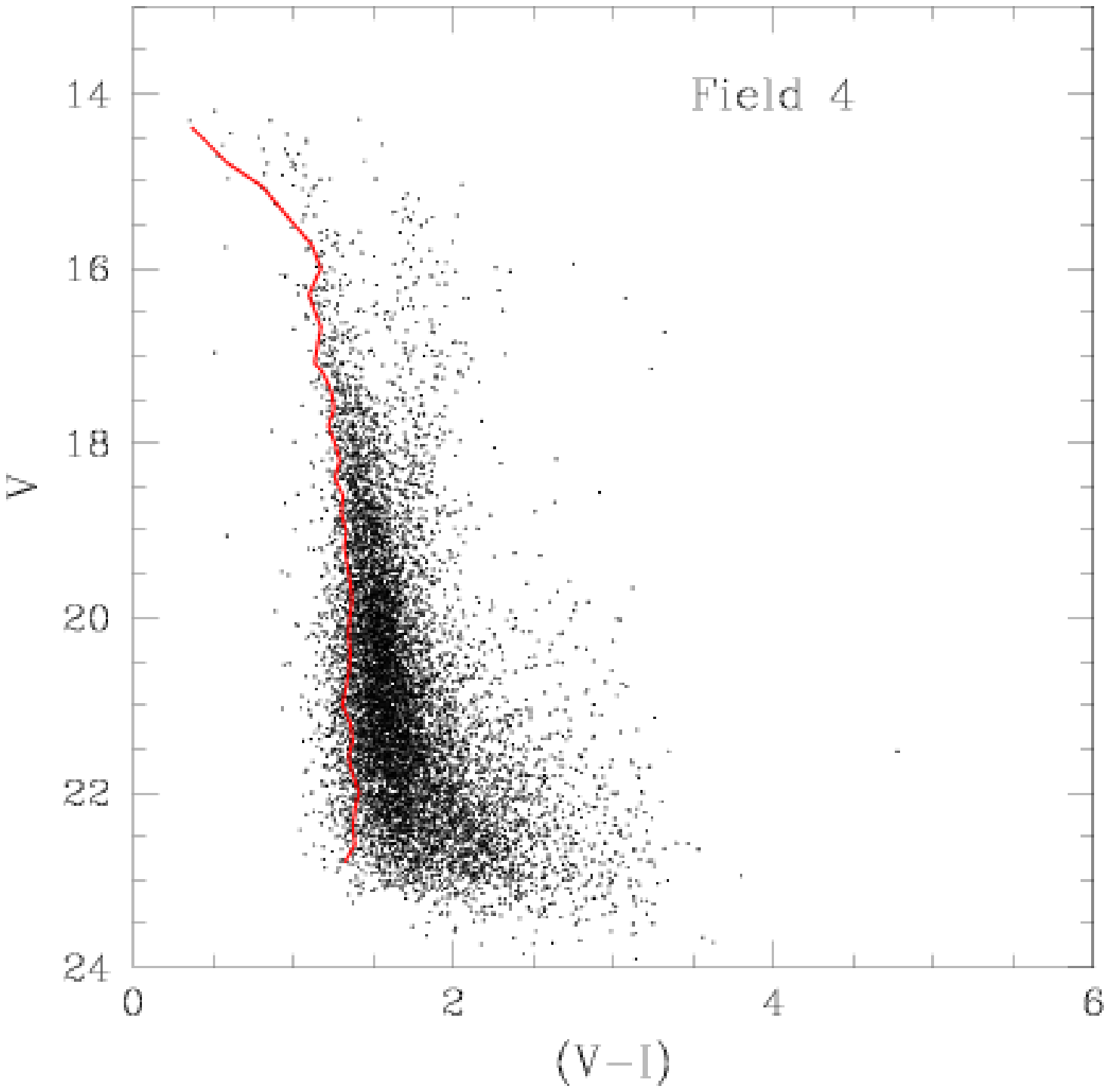}}\\
d)}
\caption{\label{CMDs} The V--(V-I) CMDs of the fields F1 (a), F2 (b), F3 (c), F4 (d). For all fields, the main sequence population
dominates the diagram. The red branch parallel to the main sequence can be
identified with evolved stars. The line shows the blue edge of main sequence (see text for details)}
\end{figure*}

In Fig. \ref{CMDs} the resulting V--(V-I)--CMDs are presented for all
the fields. 
In  these 
diagrams the main sequence population is the most prominent feature.
In F3 and F4 a red branch parallel to 
the main sequence is  visible. It is produced by   He-burning 
evolved stars distributed along the line of sight with increasing
reddening.
%% with its elongation due
%%to the increase of the reddening with the distance.
In the fields F1 and F2 this branch is scarcely populated. 
The CMDs of F3 and F4 show
a small group of stars brighter than V $\sim$ 15--15.5 on the blue side 
of the main 
sequence. In section \ref{sec_spiral}, these stars will be
shown to be 
tracers for the inner spiral arm (I).
%% They must belong to a very young population and
%%can be interepreted as a tracer for spiral arms as discussed in section 
%%\ref{sec_spiral}.

\section {Modeling the Galaxy}
\label{sec_model}
The description of the Galaxy is done with the code already described 
by Bertelli et al. (1995) and revised as described in the following sections.
First a synthetic population is generated
at varying 
 the parameters  age, metallicity range, star formation
law and initial mass function. Second, the stars are distributed along
the line of side following a model of the Galaxy, where all the components
are taken into account. Finally, the photometric 
completeness of the data is taken 
into account dividing the simulated CMD in magnitude-colour bins
and then subtracting from each bin  having N$_{th}$ stars,
  (1-$\Lambda$)N$_{th}$, where $\Lambda$ is the smallest of the V and
I completeness
factors given in Fig.\ref{comple}. 

The generation of the synthetic population makes use of the
set of stellar tracks by Girardi et al. (1996) for Z=0.0001,
Bertelli et al. (1990) for Z=0.001, Bressan et al.
(1993) for Z=0.020, Fagotto et al. (1994a,b,c) for Z=0.0004, 0.004, 0.008, 0.05,
0.10.

\subsection {The star formation rate} 
The history of the star formation in the solar neighborhood has been
derived using various methods. 
Several authors  suggest that a constant star formation rate can 
be appropriate for the disk (see among the others
Twarog 1980, Haywood et al. 1997).
On the basis of the Hipparcos data a star formation rate increasing
towards younger ages is derived (Bertelli et al 1999).
In the following,
several star formation rates going from
constant to increasing or decreasing in time are adopted and the
corresponding 
 CMDs and luminosity functions are compared with the data (see section
\ref{sec_results}).

\subsection{The position of the sun}

The position of the sun above  the Galactic disk mid-plane is found
to range from 10 to 42\,pc, the upper limit being obtained by star count
method (Stobie \& Ishida 1987). On the basis of star-counts in 12 fields in the North and South hemispheres, Humphreys \& Larsen (1995) suggest that values
as low as 20.5$\pm$3.5\,pc are more appropriate.
Even lower values of 15$-$14\,pc 
are found by Binney et al. (1997), Haywood et al. (1997), Cohen
(1995), Hammersley et al. (1995).
 However, as pointed out by Haywood et al. (1997) the expected offset of the sun deduced from star-counts
is found to depend on the scale height of the disk, in the sense that
small scale height value favors small offsets: a scale height of 350\,pc
is compatible with an offset of 20\,pc, while a scale height of 200\,pc 
suggests an offset of 15\,pc.
In the following we adopt an offset of 15\,pc. 

The distance of the sun from the Galactic center is discussed from 7 to 
8.5\,Kpc. We adopt 8\,Kpc, as a mean value.

\subsection{ Thin disk}

\subsubsection{ Mass distribution}

Two kind of mass distributions are usually adopted: 

1) a
double exponential law of the form:

\begin{equation}
$$ \rho_{disk} = \rho_0 e^{-r/h_r} e^{(z-z_\odot)/h_z} $$
\end{equation}

where $h_r$ and $h_z$ are the scale length and scale height of the disk,
respectively;

2) an exponential  distribution on the plane and sech$^2$ perpendicularly to
the plane:

\begin{equation}
$$\rho_{disk} = \rho_0 e^{-r/h_r} sech^2({z-z_\odot}/h_z) $$
\end{equation}

For z $>>$ $h_z$ the two functions are often considered equivalent.
However, in our case, this condition is not always met.
In particular, using DIRBE data,
 Freundenreich (1996) find that the sech$^2$ distribution is
greatly superior to the exponential at low latitudes ($|b| < 5^\circ$), as
in our case.
 Both distributions
will be considered and discussed.
In our formulation,
the constant $\rho_0$ can be calculated for every model imposing that the
 total number of stars in a selected region of the observational CMD is reproduced by the simulations.
From $\rho_0$ 
 the local mass density distribution will be derived (see following sections).

\subsubsection{Scale height and scale length}

The scale height $h_z$ of the thin disk varies from  325\,pc 
(Gilmore \& Reid 1983, Reid \& Majewski 1993, among others) to 200\,pc 
(Haywood et al. 1997). In our model, this 
value is assumed as a free parameter.

The scale length $h_r$ is found to vary from 3.0\,Kpc (Eaton et al. 1984,
Freudenreich 1997) 
to 2.0\,Kpc (Jones et al. 1981).
Intermediate values are derived by Ruelas-Mayorga (1991), Robin et al. (1992)
 who give $h_r$=2.5\,Kpc.

% Since the fields we are observing are too close to the Galactic plane
%and the reddening is very high,
% the models are weakly sensitive to the adopted scale length. We adopt
%3 Kpc. Other values will be discussed in the following Sections.

%The directions of the fields we are observing are more or less 
%perpendicular to the line connecting the sun and the Galactic Center.
%Hence, we are looking tangential along the disk and are therefore only 
%weakly sensitive to its scale length. We adopt 3 Kpc but other
%values will be discussed in the following sections.

In our description, $h_r$ and $h_z$ are assumed to be constant with the
Galactic radius.  This 
may not be valid in the outer Galaxy, where the dark matter might dominate
the mass. In fact radio observations suggest that the thickness of
the HI layer reaches about 400\,pc at 13\,Kpc, while the H2 layer
is considerably flatter, reaching 200\,pc at 12.5\,Kpc of galactocentric distance
(see Combes 1991 for a review).
%Simulations show that the majority of the thin/thick disk stars in the observed fields are located
%between 9--11\,kpc from the Galactic center, as a combined effect
%of the mass distribution, extinction along the line of sight and
%solid angle size. Recent models of the stellar warp
%suggest that its amplitude is increasing linearly with the distance:
%the scale height of the thin and thick disk components might flare by a factor
%5--10 between the Sun and 25\,Kpc (Gyuk et al 1999).
% However this is still matter of
%debate: Burton (1988) find that the HI warp is practically
%negligible for R $<$ 12-13\,Kpc from the center and it might be
%that the stellar disk is less warped than the gaseous component
% (Parcel et al 1997).
 However the angle
of maximum displacement above the plane is believed to range between 80$^0$(Burton 1988)
and 110$^0$ (Diplas \& Savage 1991), measured counterclockwise from the
direction l=0. The effect in the fields under discussion is negligible.

\subsubsection {The Metallicity}

 Suggestions have arisen in literature
that no age-metallicity relation is present in the thin disk. Edvardson et al. (1993)
find a spread of about 0.6 dex in $\Delta [M/H]$ among main sequence stars
of similar age in the solar neighborhood: the abundance spread for stars born at roughly the same galactocentric distance is similar in magnitude to the increase in metallicity during the lifetime of the disk.
Studies of Galactic open clusters as well as
 of B stars in the solar neighborhood came to 
the
same conclusion (Friel \& Janes 1993,
Carraro \& Chiosi 1994, Cunha \& Lambert 1992). In our simulations,
a stochastic age metallicity relation for the disk has been adopted,
with Z going from 0.008 to 0.03.

\subsubsection{Age components}

Up to now the main source of information about the age of this component
comes from the open cluster system.
If all the open clusters are member of the thin disk, a limiting
value  might be given by  NGC~6791 which is the
oldest open cluster with well-determined age. Its  age is
going from 7 to 10 Gyr (Tripicco et al. 1995).
However, Scott et al. (1995) find a somewhat peculiar kinematics for
this object. The age of Berkeley~17, which is believed to be one of the
oldest disk clusters with 12 Gyr (Phelps 1997) has recently been substantially revised
using near-IR photometry to 8-9 Gyr (Carraro et al 1999).
%The oldest age
%of metal-rich galactic disk clusters are
%8-9 Gyr (Salaris \& Weiss 1998).
A lower limit to the age of the disk is given by the white dwarfs:
 recent determinations suggest an age of 6-8 Gyr (Ruiz et al. 1996).
This value agrees quite well with 
the oldest age of field population based on Hipparcos data
 which is 11$\pm 1$ (Jimenez et al 1998). 
While the youngest age $\tau_f$ of the thin disk component
 is constrained by the CMD,
the oldest age  is assumed to be 10 Gyr.  $\tau_f$ turns
 out to be in the range 1--5 $\times 10^8$ yr in all the fields. 

The velocity dispersion of the disk stars suggest the presence of more
age components, having different scale heights (Bessel \& Stringfellow 1993).  
In the following  in addition to the old component
with large scale height, a second one is taken into account , having scale
height $h_{z1}= 100$\,pc and ages ranging from $\tau_B$ to the final age $\tau_f$.
$\tau_B$ is assumed to be  2 Gyr. 
This assumption has been verified
on the luminosity functions of the observed  fields.
When $\tau_B$ is 3 Gyr, the luminosity functions are difficultly reproduced
unless a quite high percentage of thick disk ($>$ 8\%) is  assumed.

\subsection {Thick Disc}

\subsubsection{The age and the metallicity of the thick disk}
The mean metallicity is believed to be compatible with the one of the disk globular clusters (-0.6 to -0.7), even if a metal rich tail is expected up to -0.5
together with a metal poor component
down  to -1.5 dex (Morrison et al. 1990).
Gilmore et al. (1995) find a peak of the iron  abundance distribution 
of F/G thick disk stars at -0.7 dex, with no vertical gradient.

Edvardson et al. (1993) estimate an old age for the thick disk component of
the solar neighborhood (about 12 Gyr),
 with no age gap between the thin-thick disk stars.
Gilmore et al. (1995) confirm this result. Since the abundance 
ratios of the thick disk stars reflect incorporation of iron from 
type I supernovae, these authors 
 suggest a star formation time
scale larger than the time scale of SNI which is  about 1 Gyr.
The kinematics of the thin and thick disks suggest an evolutionary connection
between the two components (Wyse \& Gilmore 1992, Twarog \& Antony-Twarog 1994). It cannot be ruled out
 that
the thick disk is the chemical precursor of the thin disk and is formed through a dissipational collapse after the halo formation and before the
end of the thin disk collapse. 

However, other scenarios are presented in literature, suggesting that the
thick disk formed after the thin disk, either for  secular kinematic
diffusion of the thin disk stars, or as the result of  a violent thin disk heating due to the
accretion of a satellite galaxy (Quinn et al. 1993, Robin et al. 1995).
At present, it is quite difficult to discriminate between these two models.
In the following, we assume the thick disk has an age range from 12 to 8 Gyr,
with constant star formation and a metallicity ranging from Z=0.0006 to 0.008.

\subsubsection{The mass distribution}

 As in the case of the thin disk,
two density laws are adopted: either double exponential, or exponential
in the plane and following sech(z/z$_\odot$) perpendicularly to the plane.

The local density of the thick disk population is poorly known, in spite of the
attempt made to measure it. From high proper motion stars, Sandage \& Fouts
(1987), Casertano et
al. (1990) find values ranging from 2\% to 10\% of the total disk density.
Large uncertainties are found also from star-counts, 
since such estimates are correlated with the determination of the scale height. Small local densities are
correlated with large scale heights. The values are going from 1\% to 10\%
of the total density. Robin et al. (1995),
Haywood et al. (1997) suggest an intermediate value of
5.6\% of the local disk population is due to thick disk. In the following we
accept only solutions compatible with a thick disk percentage between 2\% and
5\%.

\subsubsection{The scale height and scale length}

Reid \& Majewski (1993) summarized the results on the scale height of the thick disk.
Gilmore (1984) find a scale height of 1300\,pc. Norris (1987), Chen (1996),
 Gilmore \& Reid (1983) found values of 1100\,pc, 1170\,pc, and 1450\,pc 
respectively. Lower values are suggested by Gould et al. (1995) 
on the basis of HST counts, by Robin et al. (1995), Haywood et al. (1997) 
using ground based star counts: 760$\pm 50$\,pc. 

Concerning the scale length  of the thick disk  
Robin et al. (1992), Fux \& Martinet (1994),
 Robin et al. (1995) favor a value of 2.5-2.8\,kpc.
In our simulations we consider scale height and scale length of the
thick disk as free parameters.

\begin{figure*}[t]
\parbox{8.7cm}{
%\fbox{\parbox{8.0cm}{\vspace*{7.0cm} \hspace*{7cm}}}\\
%\resizebox{8.7cm}{!}{\includegraphics{f1sim1.ps}}\\
\resizebox{8.7cm}{!}{\includegraphics{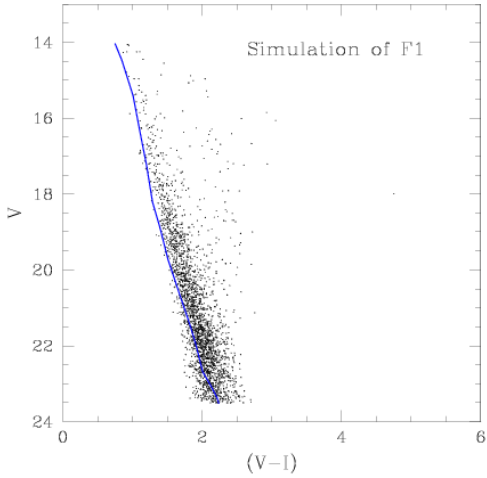}}\\
a)}
\hspace{0.5cm}
\parbox{8.7cm}{
%\fbox{\parbox{8.0cm}{\vspace*{7.0cm} \hspace*{7cm}}}\\
%\resizebox{8.7cm}{!}{\includegraphics{f2sim1.ps}}\\
\resizebox{8.7cm}{!}{\includegraphics{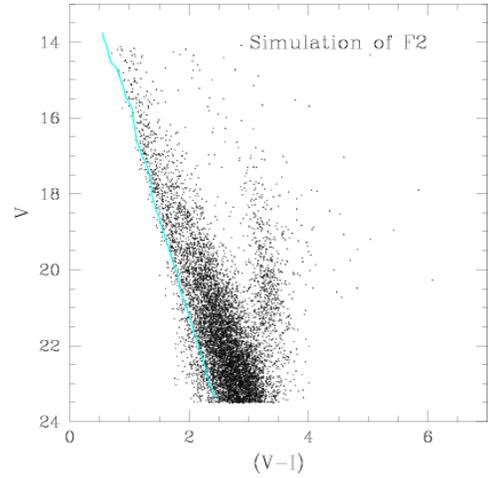}}\\
b)}
\parbox{8.7cm}{
%\resizebox{8.7cm}{!}{\includegraphics{f3sim1.ps}}\\
\resizebox{8.7cm}{!}{\includegraphics{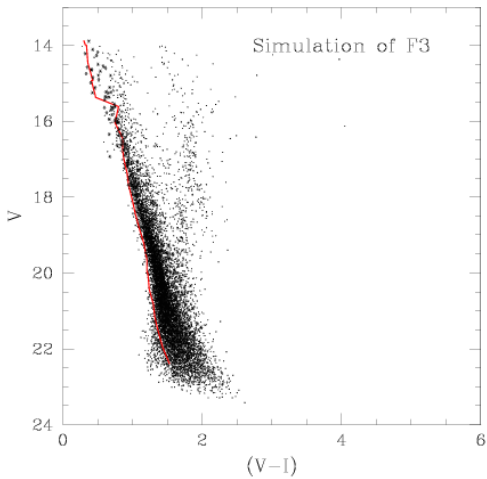}}\\
c)}
\hspace{0.5cm}
\parbox{8.7cm}{
%\resizebox{8.7cm}{!}{\includegraphics{f4sim1.ps}}\\
\resizebox{8.7cm}{!}{\includegraphics{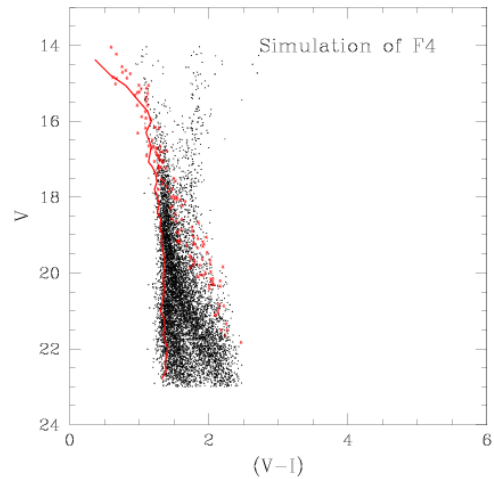}}\\
d)}
\caption{\label{CMD_simulation}The V--(V-I) CMD of the best solution for each of the fields 
is shown as example: F1 (a), F2 (b), F3 (c), F4 (d).  F1 and F2 simulations
 include  disc 
components (faint dots), while in the case of
F3 and F4   the spiral arm is as well plotted (starred dots). In each plot the line indicates the
observational edge of
the main sequence.  The star formation of the thin disc components is decreasing in time (see text for details).  }
\end{figure*}

\section {The results}
\label{sec_results}

In this Section, the fields will be analyzed separately,
deriving the reddening along the line of sight and the possible solutions.
For the best fitting solutions the simulated CMDs are displayed in Fig.\ref{CMD_simulation}. In the simulations of F3 and F4 
a young spiral-arm-like population is
also included, as described in Section~\ref{sec_spiral}.
 After discussing the single results for each field
in the following subsections, they will be compared to
derive the best fitting solution for all the fields.
Since the studied fields are  located at low Galactic latitude,
we expect them to be  more sensitive to the star formation rate
and to the scale height than to the scale lenght of the disk components.
For the same reason the parameters
of the thick disk component would not be strongly constrained.

%F3 is the closest to the Galactic plane, so we expect it to be sensitive mainly
%to the youngest population, $h_{z1}$.
% F1 is the one at higher latitude: it is most
% sensitive
%to the scale height of the thin disk, $h_z$. F2 and 
%F4 being at intermediate latitude
%can give information about the relative composition of the stellar populations,
%and about $\tau_B$.

\subsection {Discussing the CMD: Extinction determinations}

\begin{figure}[t]
%\picplace{10cm}
% \leavevmode
% \epsfysize=11cm
% \centerline{
%  \epsfbox{figred.ps}}
{\resizebox{9.0cm}{!}{\includegraphics{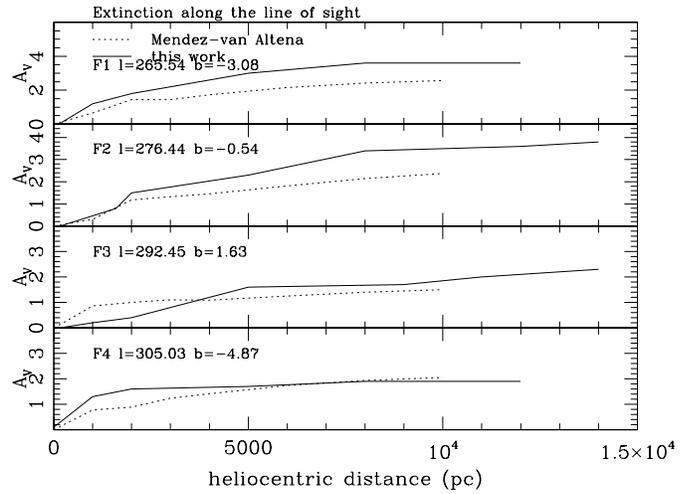}}}
%\centerline{\psfig{file=figred3.ps,height=10.truecm,width=10.truecm}}}
%\vspace*{-2cm}
\caption{\label{reddening} The reddening along the line of sight, as determined
for the four fields. For comparison, the values of \cite{Men+98} are included
in the plots.}
\end{figure}

Ng et al. (1995), Bertelli et al. (1995) proved  that the slope of the main
 sequence in the CMD of the  disk population is mainly governed by
the extinction along the line of sight. 
At each magnitude $V$ the bluest stars on the main sequence
 can be interpreted as the envelope of the main sequence turnoffs
of the  population having absolute magnitude $M_{tur}$, 
shifted towards fainter magnitudes and redder
colours
by the increasing distance and corresponding extinction.
Starting from an initial guess,
the amount of extinction at increasing  distances is adjusted
until a  satisfactory agreement between  the  main sequence
blue edge location in the data and in the theoretical simulations
is reached.
 The comparison between data and simulations 
is made using a $\chi ^2$  test.  
% we derive the values of  A$_v$ along the 
%line of sight as a function of the heliocentric distance.
The results  are given in Fig. \ref{reddening}.

We point out that F2 and F3 show  a relatively modest
  increase of the interstellar absorption along 
the line of sight ($\Delta$\.A$_v$ less than 0.5 mag)  between 1.5 and 2\,kpc distance from the Sun for F2 and between
2 and 5\,kpc for F3. Whereas this increment in F2 is probably due to 
some local density increase of the interstellar matter, the increment in F3
can be explained on larger scales since
the direction of F3  at l=292
 is crossing the inner spiral arm, if the description of
the spiral pattern given by \cite{TayCor93} is adopted. At 1.5\,Kpc distance
from the Sun, the spiral arm is reached (see section \ref{sec_spiral}),
at a height of 40\,pc above the Galactic plane. At about 5\,Kpc distance, the
spiral arm is left at a height of 150\,pc above the plane. For this field the
increase of extinction is well correlated with the intersection of the 
spiral arm pattern. In Field F4 (l=305) which does also point
towards the inner arm, no special increment of the extinction is noticeable. 
 This is not entirely 
surprising, since the the fields are chosen on the basis of their
integrated colours as having low total absorption (see Section \ref{sec_data}).

Finally, we would like to address the question  whether  the trends
in the extinction versus the distance shown in
Fig. \ref{reddening} are real.
It is quite difficult to give an estimate of the uncertainty on these
determinations. However, the simulations show that
by changing 
the extinction of $\pm 0.2$ mag at a given distance d
 a significant
shift in the location of the main sequence edge at magnitudes fainter
than $M_{tur}-5+5\times logd +A_V$ is produced.
We can safely assume that the  $A_V$ determination of Fig. \ref{reddening}
 cannot have an internal error larger  than 0.2 mag.

We point out that these determinations of extinction are dependent on the
adopted age  and metallicity range of the population.
To estimate the uncertainty due to the combined effect of different
age and metallicity distributions, we derive the extinction separately
for the three disk components, namely the young thin,
the old thin and the thick disk  described in Section \label{sec_model}.
The determination of the extinction along the line of sight $A_V$ turns out
to be different from the adopted values at maximum of 0.2.
%Since suspicion might arise that small differences in 
%the metallicity and/or star formation law can greatly influence the results,
To further check these results,
 we  make a comparison   with the values derived 
from reddening maps by \cite{Men+98}.
 Taking into account the errors on Mendez \& van Altena
determinations (0.23 mag in E$_{(B-V)}$, with A$_v$= 3.2 E$_{(B-V)}$) , 
the agreement is reasonable up to distances of about 8\,Kpc,
 at (l,b)=(292.4,1.63) (field F3), and (305.0,-4.87) (field F4).
At (l,b)=(265.54, -3.08) (field F1) and (276.4,-0.54) (field F2)
our and \cite{Men+98} A$_v$ determinations are consistent up to 3--4\,Kpc,
our values being
higher of a factor $\sim 2$ at larger distances. 

When the maximum reddening in each direction is compared with the maps
by Schlegel et al. (1998) derived using  DIRBE data,
the agreement is excellent,
in spite of the fact the authors claim their values should not be trusted
for $|b| < 5^\circ$.

%%However, though the direction l=305 is supposed to cross the inner arm at 
%%nearly the same distance as F3, for F4 the height on the plane is then about 
%%%150\,pc. If this behavior is interpreted as general, it suggests that the
%%%dark clouds in the spiral arms are stronger concentrated towards the plane 
%%%than their stellar population.
 
%Relatively modest signs of increased absorption (less than 0.5 mag) are found
%in F3 and F4 at 1.5\,Kpc,
%suggesting that
% the line of sight does not intersect dense clouds. This is not 
%surprising, since the the fields are chosen on the basis of their
%integrated colours as having low total absorption (see Section \ref{sec_data}).

\subsection {Discussing the CMD: the star formation rate}

\begin{figure}[tb]
%\picplace{10cm}
% \leavevmode
% \epsfysize=11cm
% \centerline{
%  \epsfbox{figred.ps}}
%{\resizebox{9.0cm}{!}{\includegraphics{disknuo.ps}}}
%\centerline{\psfig{file=figred2.ps,height=10.truecm,width=10.truecm}}}
%\vspace*{-2cm}\begin{figure*}[t]
\parbox{8.7cm}{
%\resizebox{8.7cm}{!}{\includegraphics{disknuoc.ps}}\\
\resizebox{8.7cm}{!}{\includegraphics{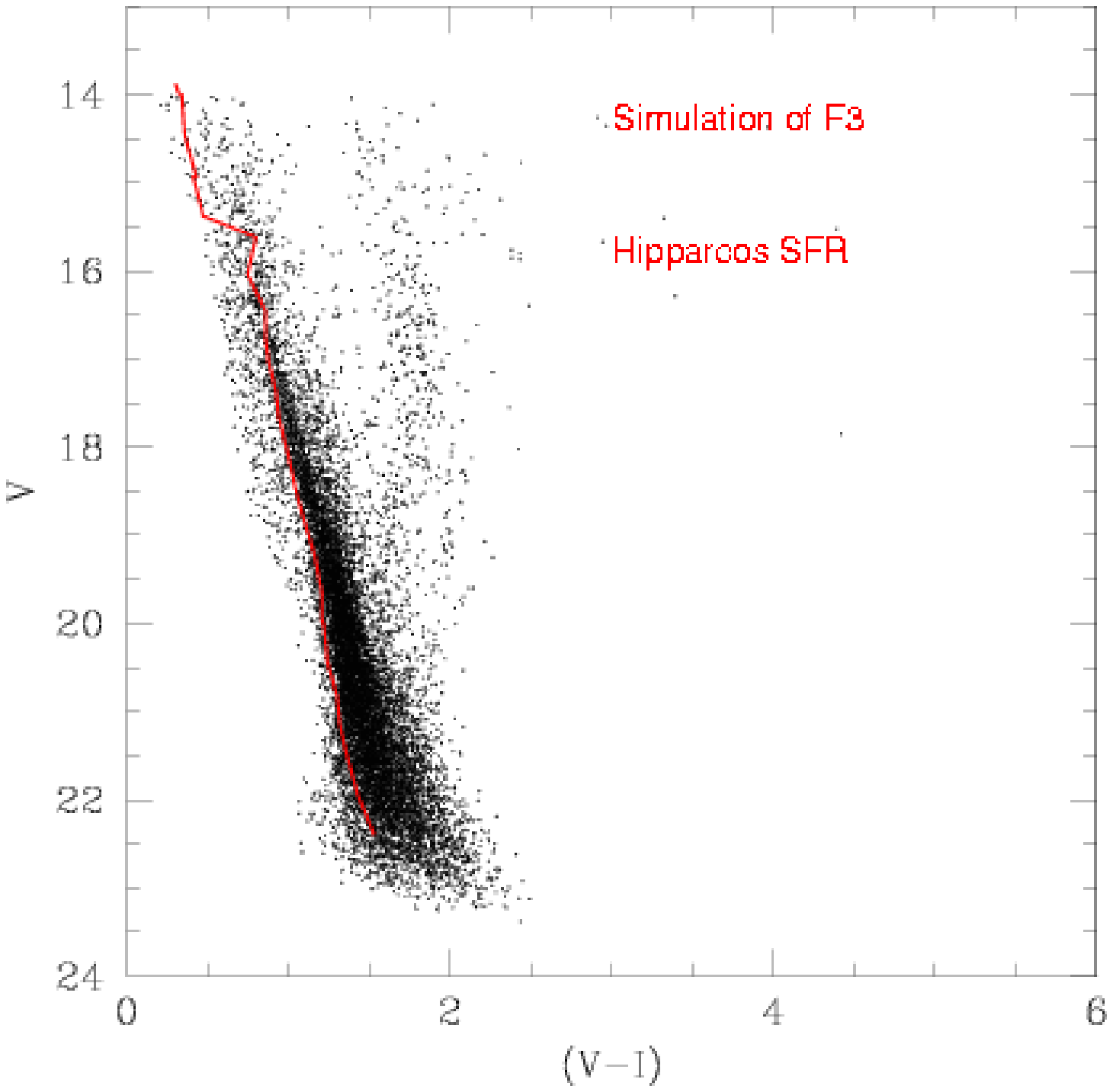}}\\
a)}
\hspace{0.5cm}
\parbox{8.7cm}{
%\resizebox{8.7cm}{!}{\includegraphics{f3n1.ps}}
\resizebox{8.7cm}{!}{\includegraphics{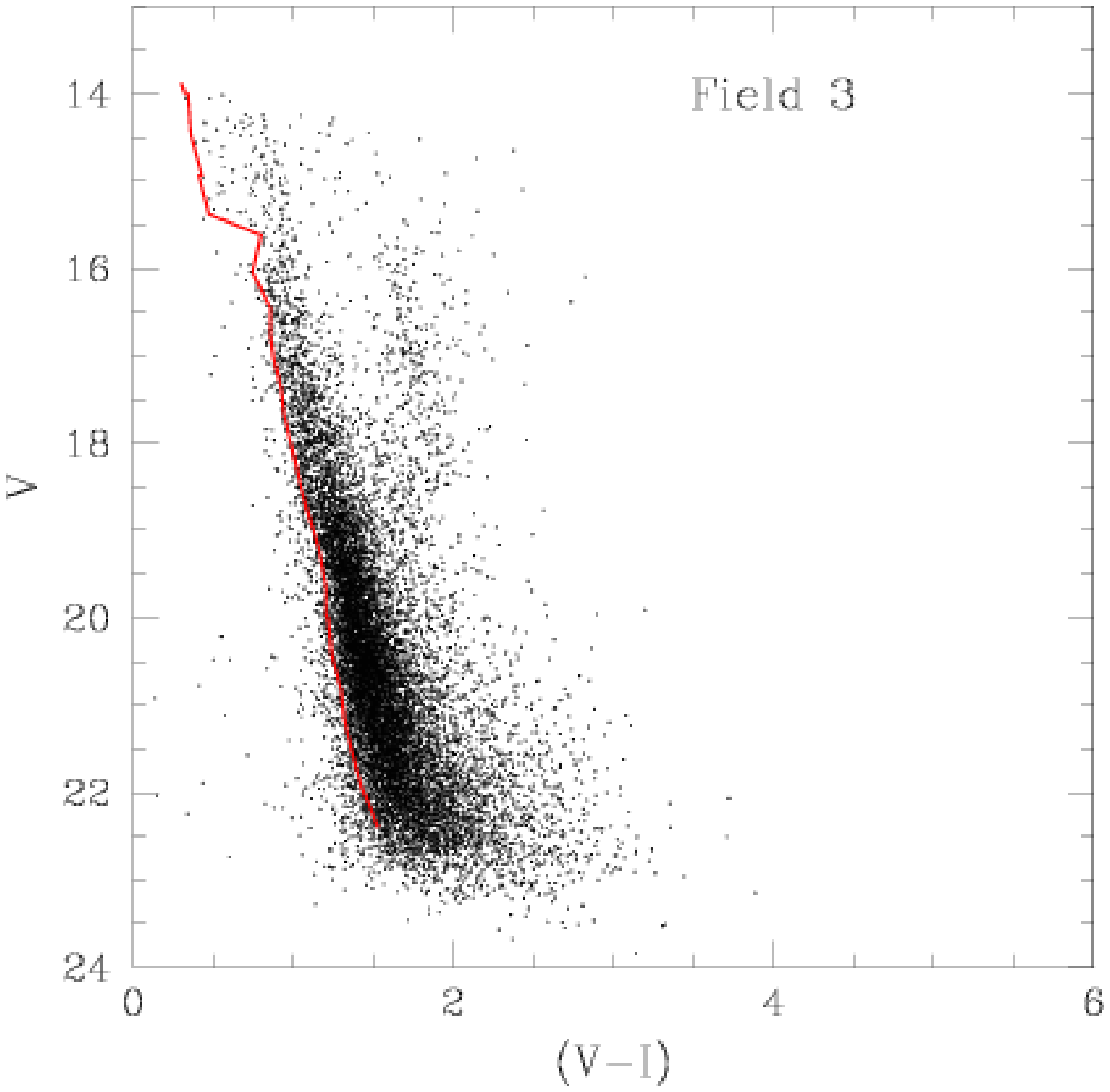}}\\
b)}

\caption{\label{disknuo} Simulation of the CMD of F3 obtained using
the Hipparcos parameterized SFR as described in the text (a). The line shows the observational 
edge of the main sequence. For sake of easy  comparison the observational
CMD of F3 (see Fig \ref{CMDs}) is as well shown (b).
}
\end{figure}

\begin{figure}[tb]
%\picplace{10cm}
% \leavevmode
% \epsfysize=11cm
% \centerline{
%  \epsfbox{figred.ps}}
\parbox{8.7cm}{
%\resizebox{8.7cm}{!}{\includegraphics{disknuof4b.ps}}\\
\resizebox{8.7cm}{!}{\includegraphics{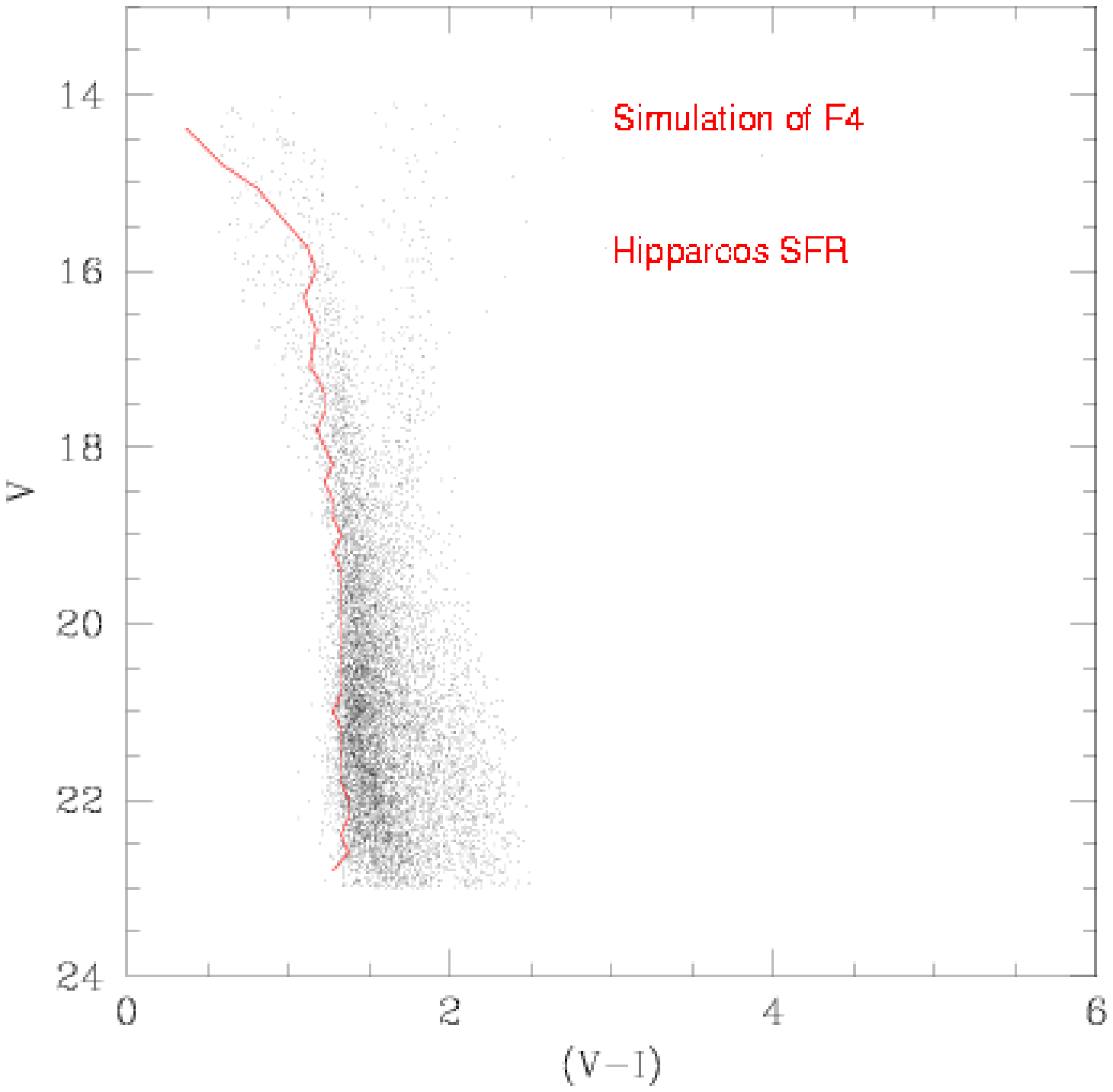}}\\
a)}
\hspace{0.5cm}
\parbox{8.7cm}{
%\resizebox{8.7cm}{!}{\includegraphics{f4n1.ps}}\\
\resizebox{8.7cm}{!}{\includegraphics{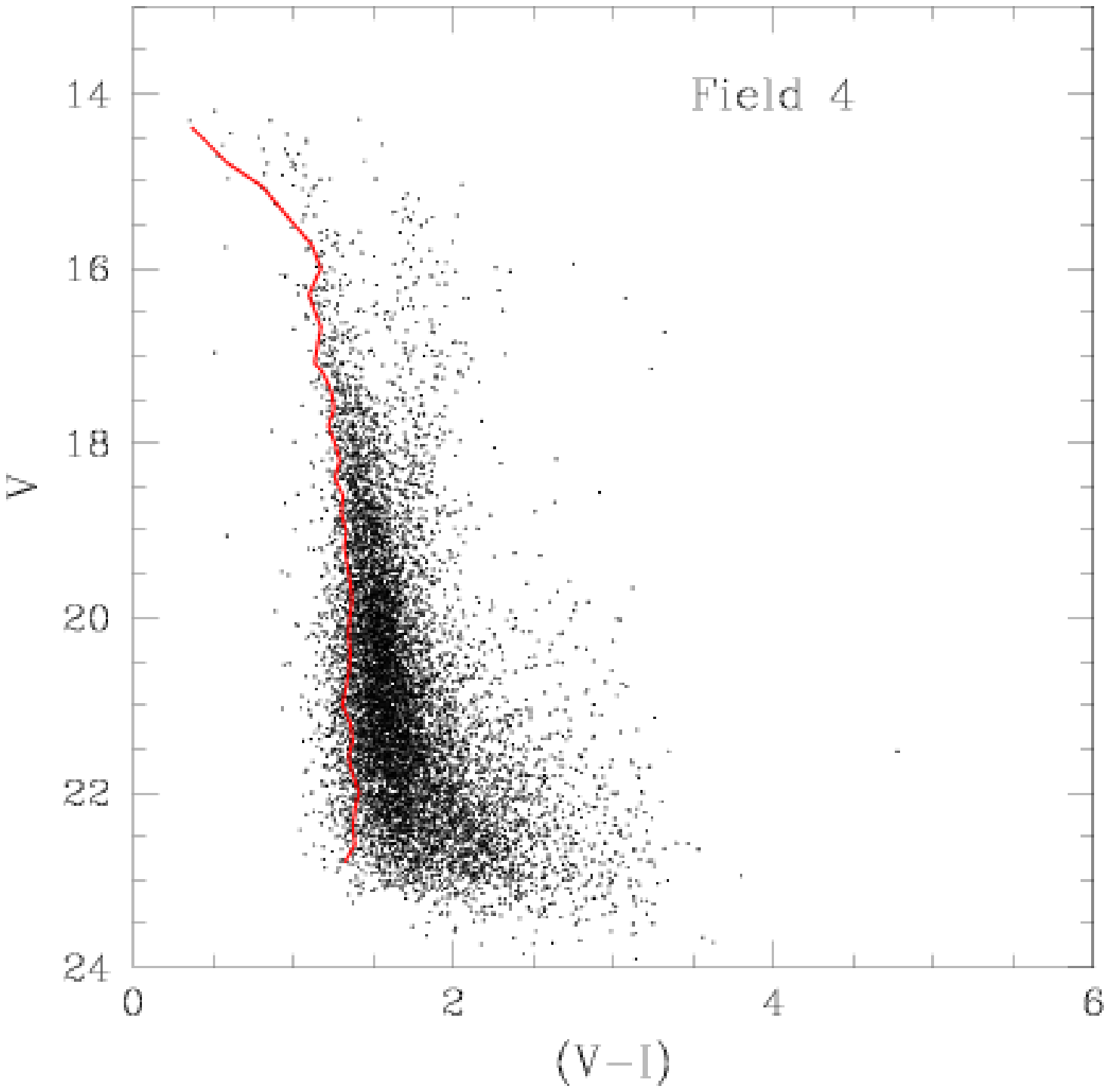}}\\
b)}
%{\resizebox{9.0cm}{!}{\includegraphics{disknuof4.ps}}}
%\centerline{\psfig{file=figred2.ps,height=10.truecm,width=10.truecm}}}
%\vspace*{-2cm}
\caption{\label{disknuof4} Simulation of the CMD of F4 obtained using
the Hipparcos parameterized SFR as described in the text (a). The line shows the observational 
edge of the main sequence. For sake of  easy  comparison the observational
CMD of F4 (see Fig \ref{CMDs}) is as well shown (b).
}
\end{figure}

\begin{figure}[tb]
%\picplace{10cm}
% \leavevmode
% \epsfysize=11cm
% \centerline{
%  \epsfbox{figred.ps}}
\parbox{8.7cm}
%{\resizebox{8.7cm}{!}{\includegraphics{disknuo_s1f1bcenc.ps}}\\
{\resizebox{8.7cm}{!}{\includegraphics{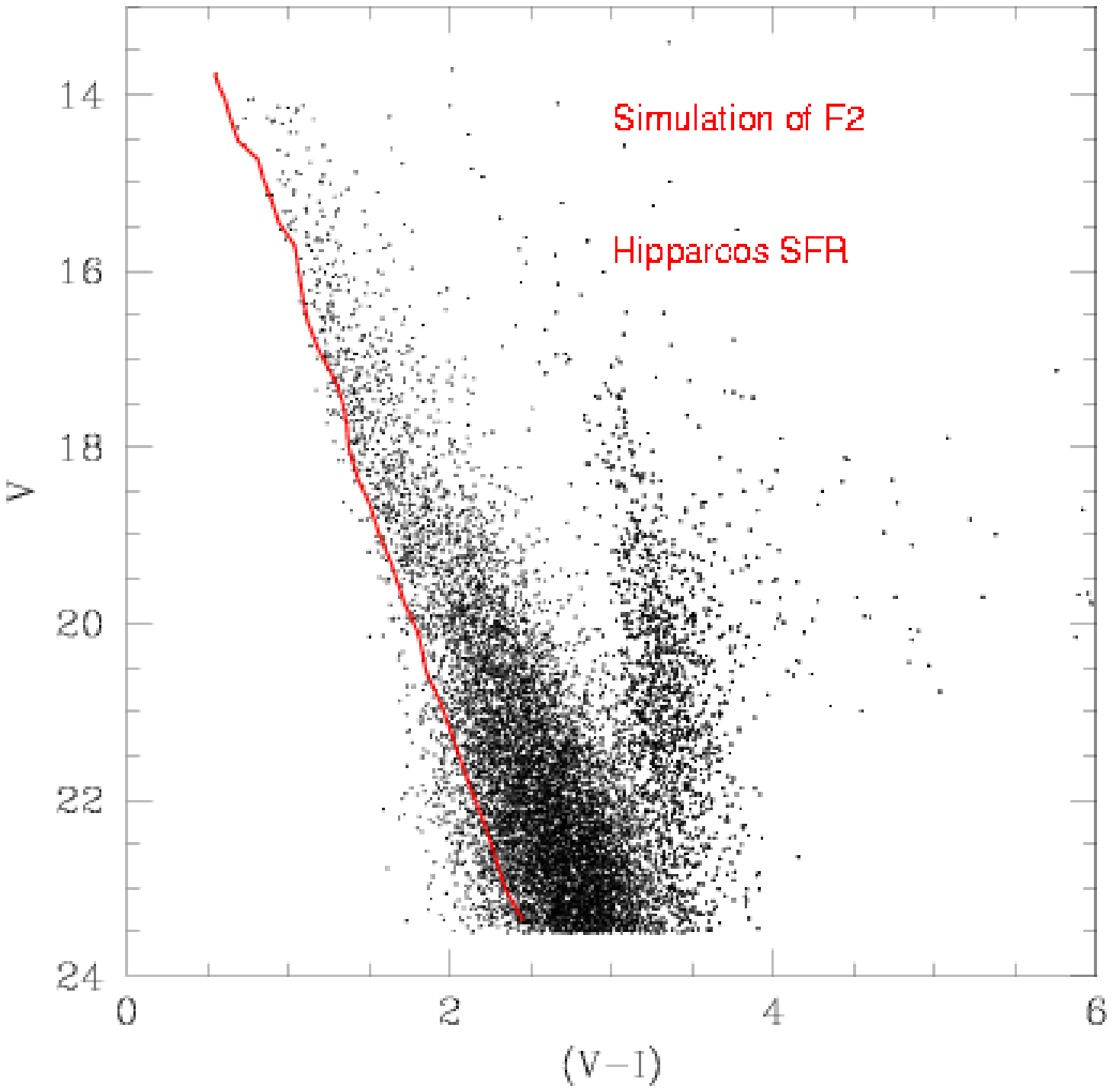}}\\
a)}
\hspace{0.5cm}
\parbox{8.7cm}{
%\resizebox{8.7cm}{!}{\includegraphics{f2n1.ps}}\\
\resizebox{8.7cm}{!}{\includegraphics{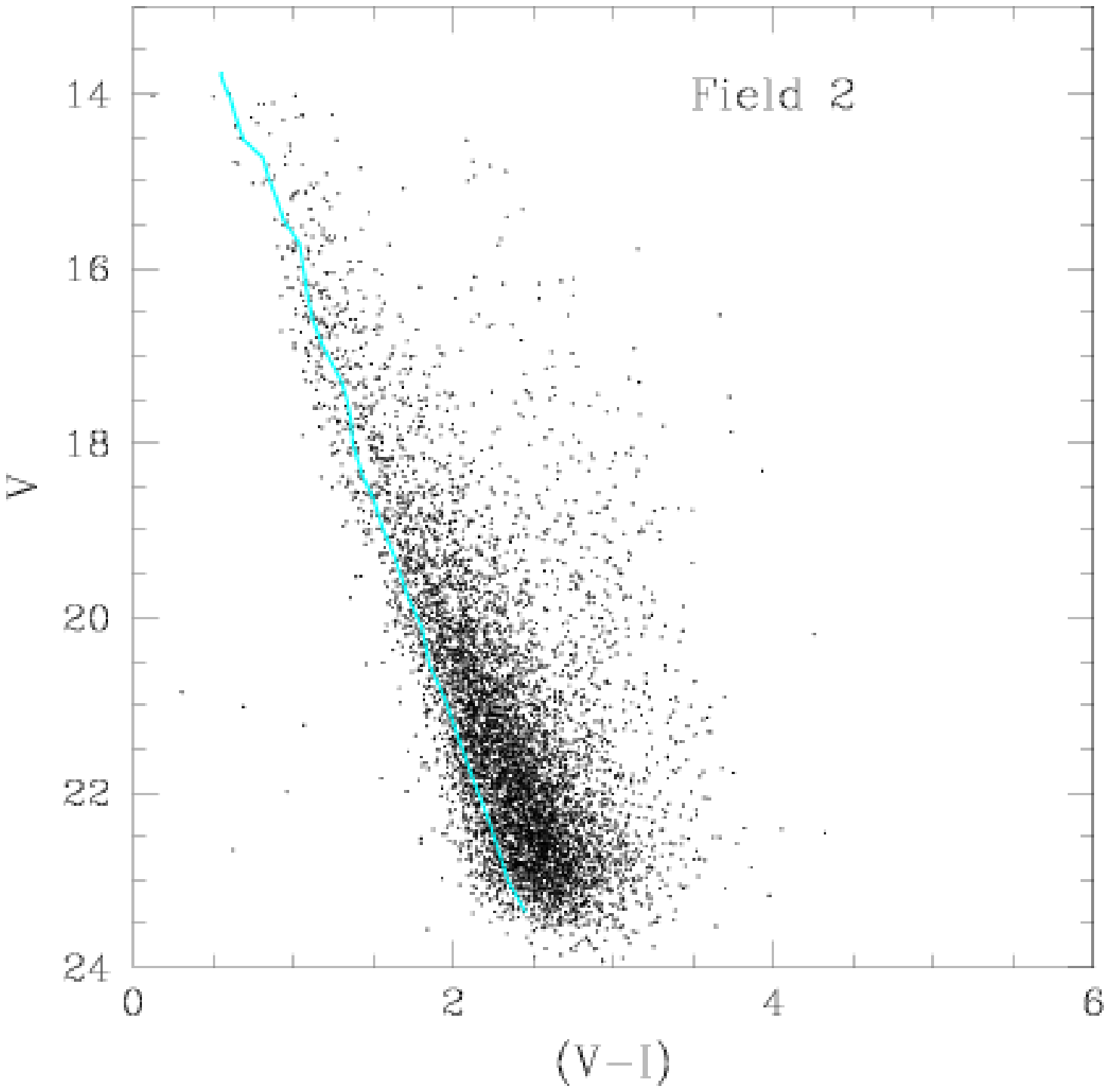}}\\
b)}
%{\resizebox{9.0cm}{!}{\includegraphics{disknuo_s1f1bcen.ps}}}
%\centerline{\psfig{file=disknuo_s1f1bcen.ps,height=10.truecm,width=10.truecm}}}
%\vspace*{-2cm}
\caption{\label{disknuo_s1f1bcen} Simulation of the CMD of F2 obtained using
the Hipparcos SFR as described in the text. The line shows the observational 
edge of the main sequence. To allow  a easy comparison, the observational
CMD of F2 (see Fig \ref{CMDs}) is as well shown (b).}
\end{figure}

\begin{figure}[tb]
%\picplace{10cm}
% \leavevmode
% \epsfysize=11cm
% \centerline{
%  \epsfbox{figred.ps}}
\parbox{8.7cm}{
%\resizebox{8.7cm}{!}{\includegraphics{fighipsim_s1f1gc.ps}}\\
\resizebox{8.7cm}{!}{\includegraphics{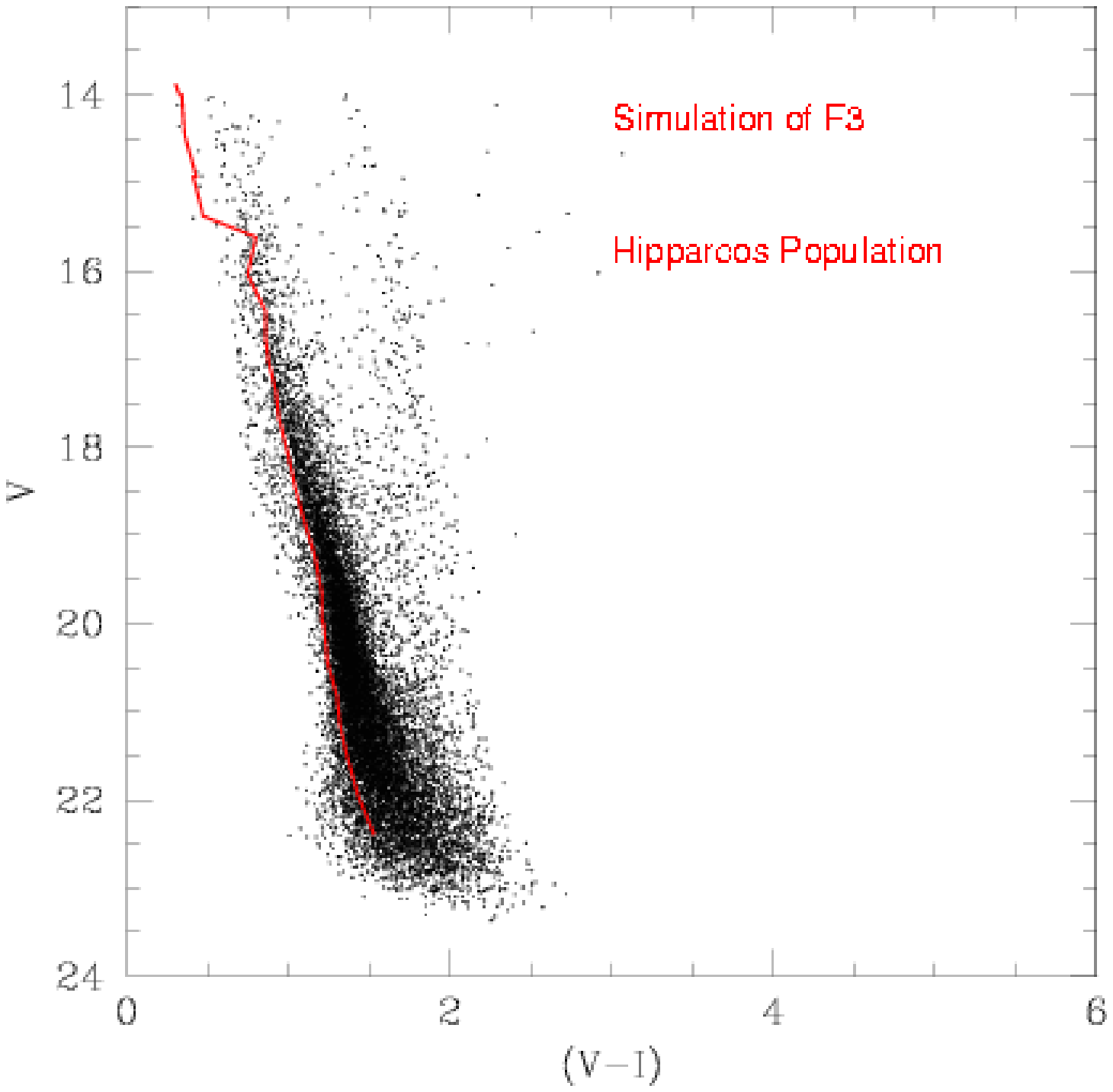}}\\
a)}
\hspace{0.5cm}
\parbox{8.7cm}{
%\resizebox{8.7cm}{!}{\includegraphics{f3n1.ps}}\\
\resizebox{8.7cm}{!}{\includegraphics{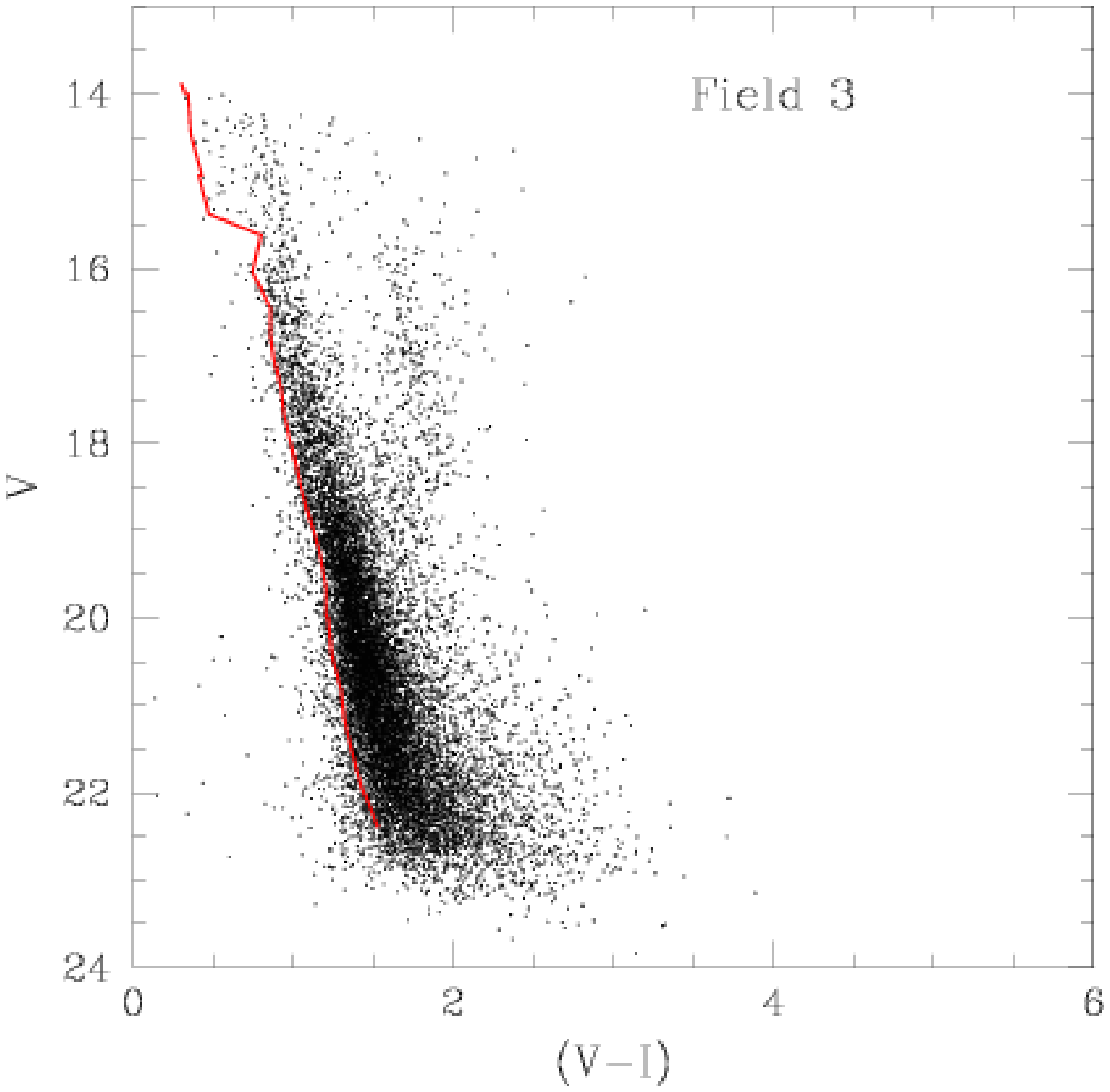}}\\
b)}
%{\resizebox{9.0cm}{!}{\includegraphics{fighipsim_s1f1g.ps}}}
%\centerline{\psfig{file=disknuo_s1f1bcen.ps,height=10.truecm,width=10.truecm}}}
%\vspace*{-2cm}
\caption{\label{hipp_s1f1g} Simulation of the CMD of F3 obtained using
the Hipparcos population as described in the text. The line shows the observational 
edge of the main sequence. For comparison   the observational
CMD of F3 (see Fig \ref{CMDs}) is  shown (b).}
\end{figure}

 To infer the star formation rate (SFR) of the thin disk component,
simulated  
CMD and LF  are calculated with constant, increasing 
or decreasing  rate and then  compared with observational  CMDs 
using a $\chi^2$ test.
%% using the shape of the HRD as 
%%indicator for the likeliness. 
The thick disc   is assumed to have a  SFR constant   
from 11\,Gyr to 8\,Gyr. Due to the low galactic latitude of the observed
fields, it is not possible to  
distinguish between constant or slightly increasing/decreasing rate
for this component. 

From the analysis of the
 Hipparcos data 
%%%allows to analyze the star formation of the solar
%%%neighborhood in great detail. 
\cite{Ber+99} derive a  disk SFR 
constant from 10\,Gyr to 4.5\,Gyr, and 
then increasing by a factor of 1.5-2
from 4.5\,Gyr to 0.1\,Gyr. 
%%It is would be reasonable to expect that
 However it is not straightforward that   
the SFR found in the solar neighborhood is  representative of
 the whole 
disk, as has already been  suggested by \cite{Ber+99}.

%Both possibilities are in contradiction to what has been found for the solar 
%neighborhood.
%From Hipparcos data, \cite{Ber+99} derived a SFR constant from 10\,Gyr to 
%4.5\,Gyr, and increasing by a factor of 1.5-2 from 4.5\,Gyr to 0.1\,Gyr*
%Therefore it seems very likely, that the SFR found in the
%solar neighborhood does not represent the whole disk, as has been 
%already suggested by \cite{Ber+99}.
%
%Suggestions have been advanced in literature that the SFR found in the 
%solar neighborhood does not represent-ate the whole disk
%(Bertelli, Bressan, Chiosi, Vallenari 1999).
% 

To assess this point, we  simulate the CMDs and luminosity functions of 
the   fields
using the parameterization of the 
SFR derived from the Hipparcos data by \cite{Ber+99}.
While the luminosity functions are not inconsistent with this SFR,
 the CMDs of F3 and F4 cannot be reproduced, since  
 too many young stars brighter than
V=19.5 and bluer than the main sequence observational edge
 are produced. This is evident
comparing the simulated and the observed
CMDs of F3 and F4 in Fig. \ref{disknuo} and Fig.\ref{disknuof4}.
The extinction cannot be responsible of this discrepancy: if the 
extinction along the line of sight at closer distances
is increased to match the observational location of the
blue edge of the main sequence at brighter magnitudes,
then  the faint main sequence turns out to be too red.

The case of F1 and F2 is substantially different. 
In these fields the blue edge of the main sequence is  reproduced  using
the Hipparcos SFR (see Fig. \ref{disknuo_s1f1bcen} for F2). However,
too many evolved stars are expected in F2 (740 stars brighter than V=22)
in comparison with the data (80 stars). No additional information is coming
from F1.
Due to the poor statistic, the observed and the
expected  number of stars in this field (63 and 74 respectively) are compatible
inside the errors.

However, it cannot be excluded that this result is dependent on the
adopted  parameterization
of the solar neighborhood star formation.
To make a further check we  use  the observed Hipparcos population,
and we distribute it along the line of sight in the disk.
The resulting CMD is presented in Fig. \ref{hipp_s1f1g} for F3. The previous conclusions are substantially
unchanged. Similar result can be reached in the case of F4, not shown for
conciseness.

%%In F1, the poor statistic does not allow for a significant distinction between 
%%the expected and the observed number of evolved stars 
%%(74 and 63 respectively), which
%%are  inside the errors.
 From this investigation we conclude
that the  solar neighborhood cannot be considered   
representative of the properties of the whole disk.
An analogous discussion can be made at varying SFR.	 
The simulations  show that any assumption of an increasing or 
even constant SFR
yields a too high number of young stars  on the blue side of the main 
sequence. Actually,
the most convincing result is obtained with a SFR constant from
10\,Gyr to 2\,Gyr,  then declining by a factor of 10 between 2\,Gyr and 
0.1\,Gyr
(see Fig.\ref{CMD_simulation}).

%%%In F3 and F4 the luminosity function of the evolved stars
%%%presents a strong decline for stars brighter than
%%%%$\rm V \approx 16$
%%%%(see Fig. \ref{CMDs}**). A decreasing SFR can reproduce this feature.
%%%%Instead   a  constant or increasing SFR produces too many bright
%%%%evolved stars.

% the CMDs
%can only be simulated if one adopts a lower limit for the age of the disk of
%$\tau_f = 0.7\rm\,Gyr$, which is a bit too old compared to the age of
%the spiral arm stars. Moreover, with this simulation the cut of evolved stars
%can not be reproduced.

As a final comment,
 the CMDs of F3 and F4 show an additional sprinkle of stars brighter 
that V $\sim$ 15.5 mag and bluer than the mean location of the main
sequence (see Fig. \ref{CMDs}). These stars cannot be reproduced unless a young burst
of star formation well confined in distance
is assumed. This feature will
be proved to be consistent with 
the presence of a spiral arm (see section \ref{sec_spiral}).

%The blue edge of the main sequence  of F1 and F2  is reproduced using
%the Hipparcos
%SFR (see Fig. \ref{disknuo_s1f1bcen} for F2). However too many evolved stars
%are expected in F2 (740 stars brighter than V=22) in comparison with the 
%data (80 stars). Due to the poor statistic, the expected and the
%observed number of stars in F1 (74 and 63 respectively) are compatible
%inside the errors.

% The best reproduction is obtained
%imposing a burst between 4 and 8 $\times 10^8$ yr having
%an intensity ratio of 2.

\subsection {The results for F3}

%%s1f1g%%
\begin{figure}[t]
%\resizebox{8.7cm}{!}{\includegraphics{lumi3.ps}}
\resizebox{8.7cm}{!}{\includegraphics{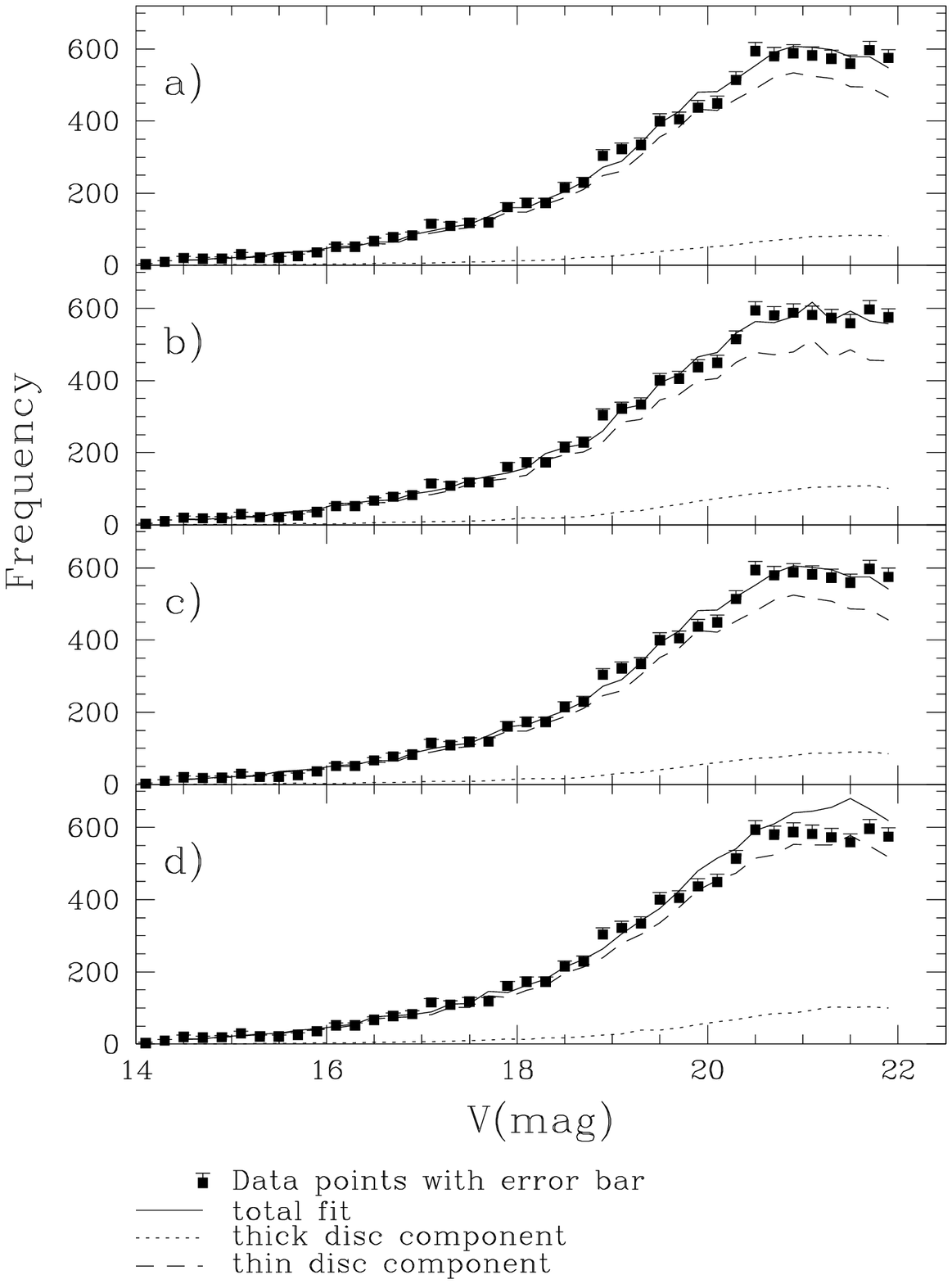}}
\caption{\label{lumi3}The luminosity function of  F3 is displayed with
the four selected solutions. Solutions a),b),c) are calculated using a sech$^2$ mass distribution, solution d) is making use of the exponential mass distribution:
\protect\\
a) $\rm h_r=2.5\,kpc$, h$_z\rm (thin)$=0.22\,kpc,  h$_z\rm (thick)$= 1.0\,kpc;
\protect\\
b) $\rm h_r=3.0\,kpc$, h$_z\rm (thin)$=0.21\,kpc, h$_z\rm (thick)$= 1.5\,kpc;
\protect\\
c) $\rm h_r=3.0\,kpc$, h$_z\rm (thin)$=0.22\,kpc,  h$_z\rm (thick)$= 1.0\,kpc;
\protect\\
d) $\rm h_r=3.0\,kpc$, h$_z\rm (thin)$=0.20\,kpc,  h$_z\rm (thick)$= 1.0\,kpc.}
\end{figure}

\begin{table}[t!]
\caption{\label{solf3} Solutions for F3}
\begin{minipage}[t]{8cm}
Density law : sech$^2$\\
\begin{tabular}{ c | c c | c  | c }
\hline
 \noalign{\smallskip}
 \multicolumn{1}{c}{} & \multicolumn{2}{c}{thick disc}&\multicolumn{1}{c}{thin disc}&\\
\hline
$h_r$/kpc & $h_z$/kpc & $\frac{m_{thick}}{m_{tot}}$ & $h_z$/kpc & $\chi ^2$\\
\hline
\multicolumn{5}{c}{ Maximum disc radius = 18 Kpc}\\
\hline
%mit evolved stars--cut a 2 gyr
%3.0 & 1.2 & 1.3\% & 0.30  & 100 \\
%3.0 & 1.2 & 1.3\% & 0.25 & 120 \\
2.0 & 0.7 & 3.0\% & 0.23$\pm 0.03$& 1.1 \\ %34 \\
2.0 & 1.5 & 2.0\% & 0.23$\pm 0.03$& 1.2 \\ %36 \\
2.5 & 0.7& 2.5\%  & 0.22$\pm 0.03$& 1.1 \\ %34\\
2.5 & 1.0 & 2.0\% & 0.21$\pm 0.02$& 1.2 \\ %35 \\
2.5 & 1.5 & 2.0\% & 0.22$\pm 0.03$& 1.3 \\ %37 \\
3.0 & 1.0 & 2.0\% & 0.22$\pm 0.03$& 1.3 \\ %38 \\
3.0 & 1.2 & 1.3\% & 0.22$\pm 0.03$& 1.2 \\ %35 \\
3.0 & 1.5 & 1.3\% & 0.21$\pm 0.03$& 1.1 \\ %33 \\
\hline
\multicolumn{5}{c}{ Maximum disc radius = 14\,Kpc}\\
\hline
2.0 & 1.5 & 2.5\% & 0.22$\pm 0.02$ & 1.4 \\ %41\\
2.5 & 1.5 & 2.2\% & 0.21$\pm 0.01$ & 1.3 \\ %38\\
3.0 & 1.0 & 2.3\% & 0.21$\pm 0.01$ & 1.2 \\ %36\\
\hline
\end{tabular}
\end{minipage}
\\[0.6cm]
\begin{minipage}[t]{8cm}
Density law : double exponential\\
\begin{tabular}{ c | c c | c  | c }
\hline
 \noalign{\smallskip}
 \multicolumn{1}{c}{} & \multicolumn{2}{c}{thick disc}&\multicolumn{1}{c}{thin disc}&\\
\hline
$h_r$/kpc & $h_z$/kpc & $\frac{m_{thick}}{m_{tot}}$ & $h_z$/kpc & $\chi ^2$\\
\hline
\multicolumn{5}{c}{ Maximum disc radius = 18\,Kpc}\\
\hline
2.0 & 1.0 & 1.5\% & 0.30$\pm 0.01$ & 1.6 \\ %47\\
2.0 & 1.5 & 1.3\% & 0.22$\pm 0.02$ & 1.7 \\ %50\\
2.5 & 1.0 & 1.3\% & 0.25$\pm 0.01$ & 1.9 \\ %56\\
2.5 & 1.0 & 2.0\% & 0.20$\pm 0.01$ & 2.2 \\ %65\\
2.5 & 1.5 & 1.2\% & 0.20$\pm 0.01$ & 2.0 \\ %59\\
3.0 & 1.0 & 1.3\% & 0.20$\pm 0.01$ & 1.9 \\ %56\\
\hline
\end{tabular}
\end{minipage}
\end{table}

%In view of the uncertainties described in the previous sections,
Various combinations of the disk parameters, at changing
 scale height and scale length of the thick disk,  scale
length of the thin disk. Both spatial
distributions (sech$^2$ and double exponential) for the disk are tested.
The observational  luminosity functions  are compared with the
simulations, using a $\chi$ square test.
Fig.\ref{lumi3} presents the comparison of the observational luminosity function
with four simulations.
%Finally we impose that the percentage in mass of the thick disk 
%to the thin disk at the solar
%neighborhood is between 3\% and 5\%. The sensitivity of the results to the
%adopted percentage is tested.
%The $\chi$ square test is used to discuss the goodness of the fit.

For every set of models the most convincing solutions
for 
%% both constant and
 decreasing stars formation 
 are listed in Table \ref{solf3} together with
the  $\chi^2 = 1/\nu \sum _i (a_i-b_i)^2/a_i$ where $a_i$ and $b_i$
are the observed and the expected number of stars per magnitude bin,
and $\nu$ is the number of degrees of freedom (which is equal to the number
of bins minus 1, since the in the simulations we impose that the
total number of model stars is equal to the observed total number of stars)

%As the solutions for the decreasing SFR are similar to the
%ones derived for the constant SFR, but for a slightly lower percentage
%of thick disk stars, we have not listed all of them again.

%%Due to the location of the field close to the Galactic plane,
%%the solutions are insensitive to the thich disk  h$_r$.
%%%%%%%%%%%%%%%%%%%%%%%%%%%%%%%%%%%%%%%%%%%%%%%%%%%%%%%%%%%%%%%%%%%%%%%%%%%%%%%

%higher percentage of
%the thick disk component.
As expected,
the scale height of the thick disk h$_z$  is poorly constrained
by the data.  Any value of h$_z$  in the range 700--1500\,pc, turns out
to be acceptable,  higher values resulting
 in a  slightly lower percentage of the 
thick disk component. The most convincing fits are obtained for
a thick disk percentage of 2-3\%. 
The sech$^2$ mass distribution
seems to be favored by the data, resulting in more convincing 
luminosity function fits. Values of thin disk h$_z$ of 220$\pm$30 pc
seem to be more consistent with the data. 

% Acceptable solutions are going from 200 to 350\,pc.
%For scale heights larger than 330\,pc however, the percentage of the thick disk
%component can hardly be estimated, as all values between 0.1\% and 2\% seem to
%be possible. Also the simulation of the luminosity function is a bit poor, 
%resulting in the high $\chi$-value.  

%Using a decreasing SFR that better reproduces the features of the CMD,

%With such a percentage of thick disk the double exponential distribution
%seems to be slightly favored. It results in smaller scale height of
%the thin disk. A higher scale length instead favors smaller scale height
%values. 
%While almost all the solutions suggests small values of the scale height
%(less than 250 pc), the best fit is obtained for thick disk
%sl=2000 pc, sh=1200 pc,
%resulting in a thin disk scale height 270$\pm 10$ pc.
Imposing a cut of the disk at 14\,Kpc, the fit is not substantially improved, 
as is expected due to the low Galactic latitude.
% and  a smaller scale height is derived.
%(sh=200-250).  
%We notice however that more convincing fits are obtained using 2\% of thick disk stars.
%Finally small scale length value ($\sim 2$ Kpc) seems to be favored.

\subsection { The results for F4 }

%% s3f4

\begin{figure}[bt]
%\resizebox{8.7cm}{!}{\includegraphics{lumi4.ps}}
\resizebox{8.7cm}{!}{\includegraphics{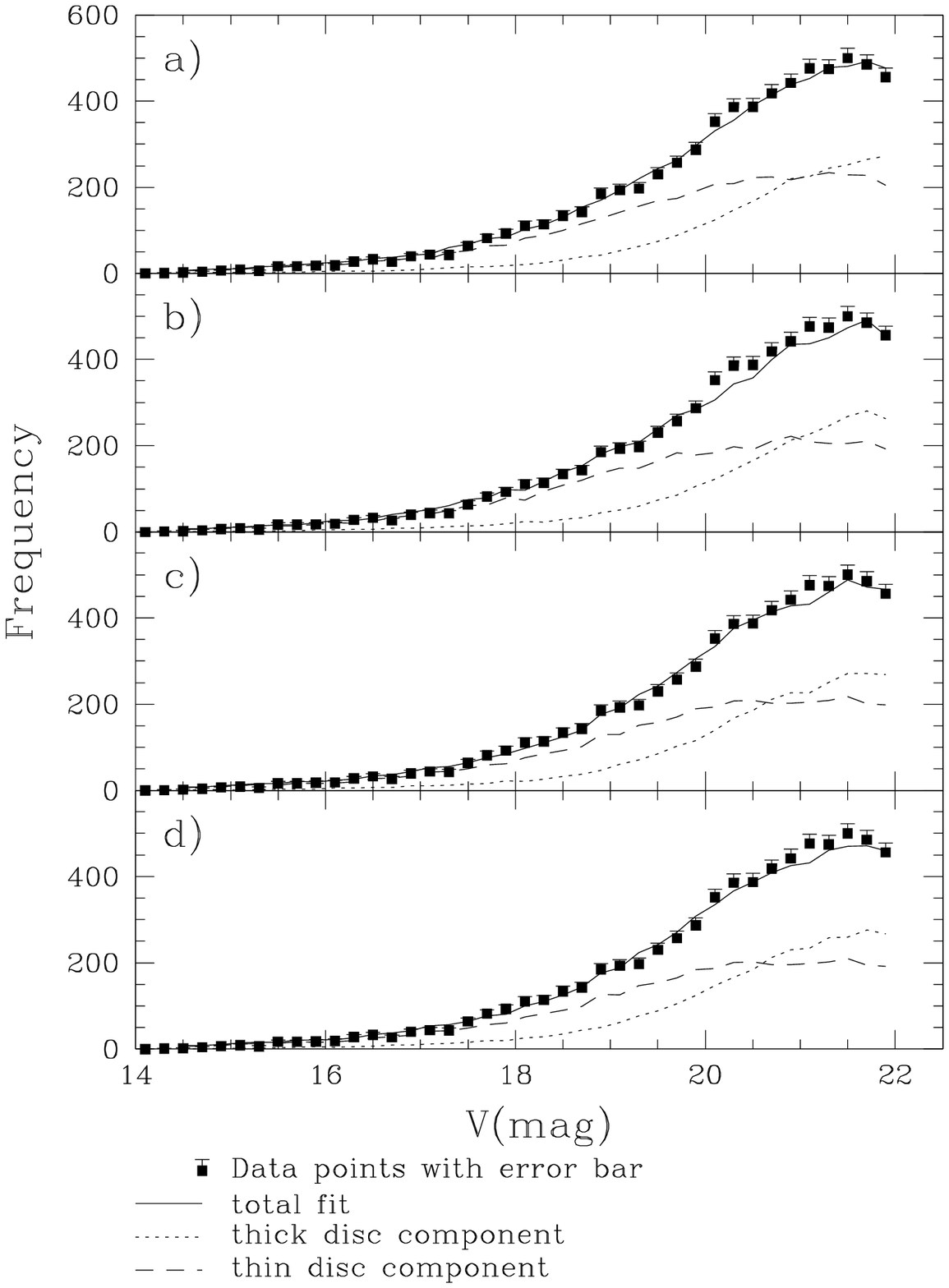}}
\caption{\label{lumi4}The luminosity function of Field 4 is displayed with
  four selected solutions for sech$^2$ mass distribution: \protect\\
a) $\rm h_r=3.0\,kpc$, h$_z\rm (thin)=0.25\,kpc$, h$_r\rm (thick)=1.5\,kpc$;
\protect\\
b) $\rm h_r=2.0\,kpc$, h$_z\rm (thin)=0.28\,kpc$, h$_r\rm (thick)=1.1\,kpc$;
\protect\\
c) $\rm h_r=2.5\,kpc$, h$_z\rm (thin)=0.28\,kpc$, h$_r\rm (thick)=1.5\,kpc$;
\protect\\
d) $\rm h_r=3.0\,kpc$, h$_z\rm (thin)=0.27\,kpc$, h$_r\rm (thick)=0.8\,kpc$.}
\end{figure}

\begin{table}[t!]                                                              
\caption{\label{solf4} Solutions for F4} 
\begin{minipage}[t]{8cm}
Density law : sech$^2$\\
\begin{tabular}{ c | c c | c  | c }
\hline
 \noalign{\smallskip}
 \multicolumn{1}{c}{} & \multicolumn{2}{c}{thick disc}&\multicolumn{1}{c}{thin disc}&\\
\hline
$h_r$/kpc & $h_z$/kpc & $\frac{m_{thick}}{m_{tot}}$ & $h_z$/kpc & $\chi ^2$\\
\hline
\multicolumn{5}{c}{ Maximum disc radius = 18\,Kpc}\\
\hline
%cut old/young bei 2Gyr
2.0 & 1.1 & 4\% & 0.28$\pm 0.03$ & 1.2 \\ %32 \\
%%2.0 & 1.5 & 5.0\% & 0.35$\pm 0.02$ & 0.9 \\ %35\\
2.0 & 1.5 & 4\% & 0.28$\pm 0.03$ & 0.9 \\ %26 \\
%%2.5 & 1.2 & 3.0\% & 0.35$\pm 0.02$ & 1.1 \\ %23 \\
%%2.5 & 1.2 & 3.3\% & 0.30$\pm 0.02$ & 24 \\
%%2.5 & 1.2 & 3.1\% & 0.32$\pm 0.03$ & 24 \\
2.5 & 1.5 & 4\% & 0.28$\pm 0.03$ & 0.9 \\ %26 \\

%3.0 & 1.2 & 4.6\% & 0.30$\pm 0.02$ & 29 \\
%3.0 & 1.5 & 5.0\% & 0.25$\pm 0.02$ & 51 \\
3.0 & 0.8 & 4.\% & 0.27$\pm 0.02$ & 1.1 \\ %31 \\
3.0 & 1.5 & 3.\% & 0.25$\pm 0.03$ & 0.7 \\ %20 \\
\hline 
\end{tabular}
\end{minipage}
\\[0.6cm]
\begin{minipage}[t]{8cm}
Density law :  double exponential\\
\begin{tabular}{ c | c c | c  | c }
\hline
 \noalign{\smallskip}
 \multicolumn{1}{c}{} & \multicolumn{2}{c}{thick disc}&\multicolumn{1}{c}{thin disc}&\\
\hline
$h_r$/kpc & $h_z$/kpc & $\frac{m_{thick}}{m_{tot}}$ & $h_z$/kpc & $\chi ^2$\\
\hline
\multicolumn{5}{c}{ Maximum disk radius = 18\,Kpc}\\
\hline
%2.0 & 1.2 & 4.0\% & 0.30 & 30\\
2.0 & 1.2 & 3.0\% & 0.25 $\pm$ 0.02 & 1.0 \\ %26\\
2.0 & 1.5 & 4.0\% & 0.21 $\pm$ 0.01 & 0.8 \\ %21\\
2.5 & 1.5 & 3.2\% & 0.21$\pm$ 0.01 & 1.1 \\ %32\\ 
3.0 & 1.5 & 3.0\% & 0.21$\pm$ 0.01 & 1.1 \\ %30\\
%3.0 & 1.2 & 3.7\% & 0.25 & 29\\
%3.0 & 1.2 & 4.1\% & 0.20 & 32\\
\hline
\end{tabular}
\end{minipage}

\end{table}

Table \ref{solf4} presents the solutions and 
Fig.\ref{lumi4} shows the comparison of the observational luminosity function
with four simulations.
%Thin disk scale height larger than 330 pc are obtained with sech$^2$
%distribution.
Analogously to F3, the sech$^2$ mass 
distribution is slightly favored, again resulting in higher 
percentages of the thick disk component. 
Compared to Field 3, the percentage of thick disk is generally higher in this
field: 
%Particularly critical is the percentage of thick disk:  
to reproduce the data with constant SFR at least
3--4\% of thick disk is needed.
The data are not sensitive to the scale lenght and scale height  of the thick
disk. 

%Slightly lower values are possible by using the decreasing SFR.
%, the sh is coming out to be around 300$\pm 20$ pc. It is necessary to
%impose at least 6\% of the  thick disk to obtain a sh as low as 250 pc,
%what is however the best fit.
%Imposing a cut of the disk at 14 Kpc, a good fit is obtained again
%with sh=300 and double exponential distribution.
Concerning the scale height of the think disk, the most convincing solutions
are for  270 $\pm$ 30 pc.
Scale lenght values as low as 1.1\,Kpc result is a less good fit of
the luminosity function,  while all the values in the range 1.2--1.5\,Kpc
are consistent with the data.  

%Due to its larger distance to the Galactic plane, F4 seems to be more 
%sensitive to the scale height of the thin disk than F3. 
%Values below 250\,pc do
%not yield convincent solutions. For these low values, 
%the fit can be improved by increasing the percentage of the thick disk 
%component, but even with a percentage as high as 10\% the luminosity function 
%is not properly reproduced. 
%For the thick disk parameters, F4 is as unsensitive as F3.
%All values for h$_z$ between 800 and 1500\,pc are possible,
%the higher values resulting in generally lower percentages of the
%thick disk component.

Analogously to F3, introducing a cut of the disk at 14\,Kpc, we do not improve the fit 
of the luminosity function.

%\subsection { The results for F1 and F2}
%%s1f1j and s1f1b

%\begin{table}[tb]
%\caption{\label{solf1} Solutions for F1}
%\begin{tabular}{ c c c | c  | c }
%\%hline
% \noalign{\smallskip}
% \multicolumn{3}{c}{thick disk}&\multicolumn{1}{c}{thin disk}&$\chi ^2$\\
%\hline
%$h_z$\,(kpc) & $h_r$\,(kpc) & share & $h_z$\,(kpc)&\\
%\hline
%\multicolumn{5}{c}{Density law : sech$^2$}\\
%\hline 
%\multicolumn{5}{c}{ Maximum disk radius = 18\,Kpc}\\
%\hline
%1.2&3.0&3\% & 0.270$\pm 0.20$&22\\
%1.2&2.5&4\% & 0.290$\pm 0.20$&23\\
%1.2&2.0&5\% & 0.270$\pm 0.20$&40\\
%\hline 
%\hline
%\multicolumn{5}{c}{Density law : double exponential}\\
%\hline
%\multicolumn{5}{c}{ Maximum disk radius = 18\,Kpc}\\
%\hline
%1.2&3.0&3\% & 0.270$\pm 0.20$&22\\
%1.2&2.5&4\%&0.250$\pm 0.20$&23\\
%1.2&2.0&4\%&* $\pm 0.20$&42\\
%\hline 
%\end{tabular}
%\end{table}

%Table \ref{solf1} gives the results for F1. The  most convincing solutions
%are for  h$_z$=280$\pm 30$\,pc and h$_r$=2750$\pm$250\,pc.
%The percentage of thick disk is about 3--4\%.

\subsection{Hunting for common solutions}

The scale height of the thin disk derived for F4 are in the mean higher
than the ones derived for F3. However they are consistent at the 2-$\sigma$
level, if using the best value of 280$\pm$30 pc for F4 and 220$\pm$30
for F3. The most convincing
common solution with  decreasing star formation rate is found 
 about 
2--4\% of thick disk, a scale height around $h_z = 250 \pm 60\rm\,pc$
and the sech$^2$ distribution.

Due to the poor statistics, F1 and F2  (see Table \ref{coordinates})
do not set further constraints on the determination of the 
scale height and lenght of the disk. 
The errors are relatively large and all
the solutions, we found for the fields F3 and F4 do also fit the fields F1 and 
F2. Hence we can only conclude, that F1 and F2 are consistent within the 
uncertainties with the other fields.

%Using the double exponential distribution yields the same solution but with 
%a percentage of 3-4\% of thick disk.

\subsection{ Discussing the mass distribution}

We use the value of the  central value of the mass distribution
 $\rho_0$ (see Eqs.2 and 3)
 to derive the local mass 
density.
$\rho_0$ can be derived for each field
imposing the total number of stars in a
selected region of the CMD.
%Simple calculations allow to verify that $\rho_0$ can express the
%central value of the mass density in stars more massive than 0.6 M$\odot$
So, we expect that 
inhomogeneities of the mass distribution reflect  in  different
constants in different fields.
%For common best solution,
%the mass constants are quite similar for the thick disk, 
%(4e7 M$\odot$/pc$^3$**)
% and the old
%component of the thin disk. 
%(7.8e8 M$\odot$/pc$^3$**).

Taking into account all the disk components, there is a slight evidence
that the total mass density might be higher of a factor  1.5-2  in F4 and F1
than in F2 and F3,
the main difference residing in the mass of the component
older than 2 Gyr. 
However it is
not clear whether  this effect is real then reflecting inhomogeneities
of the disk on small scale, or  it must simply be interpreted as due to
the uncertainty on the mass determination.
From this constant, we can derive the mass density in the local
neighborhood  in stars more massive than  0.1 M$_\odot$,
calculated using Kroupa (2000) IMF.
 For the best solutions, we derive a total local star
density of  0.025 M$_\odot$ pc $^{-3}$ in F3 and 0.036 in F4, 0.034 in F1 and 
 0.022 in F2. These results are in agreement with previous determinations
of the local density.
Oort (1960) find 0.18 M$_\odot$ pc $^{-3}$ as total local mass density,
 where the total mass density in
stars and interstellar matter would not exceed 0.08  M$_\odot$ pc $^{-3}$.
Lower values are derived by
Creze et al (1998) who estimate that  the local mass density in 
stars  can be 0.042-0.048  M$_\odot$ pc $^{-3}$,
 including 0.015 in stellar remnants, while
the total dynamical mass density cannot exceed 0.065-0.10   
M$_\odot$ pc $^{-3}$.

\begin{figure}[t]
%\rotatebox{-90}{\resizebox{!}{8.5cm}{\includegraphics{aufbau.ps}}}
\rotatebox{-90}{\resizebox{!}{8.5cm}{\includegraphics{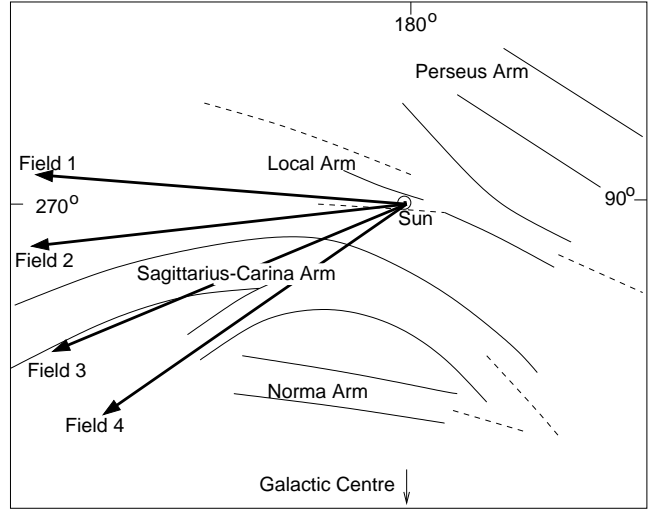}}}
\caption{\label{aufbau} The position of the four fields with respect
to the Galactic structure (\cite{Hum76}). 
F3 and F4 point towards the Sgr-Car arm, 
whereas no spiral arm is present in the direction of F1 and F2.}
\end{figure}

\section {Discussing the presence of a spiral arm }
\label{sec_spiral}

As already pointed out in the previous sections,
the CMDs of fields F3 and F4 (see Fig. \ref{CMDs}) 
show the presence of 
 a few 
stars  brighter than V $\sim 15--15.5$ mag for F4 and F3 respectively  and bluer than the average main sequence.
 Stars in this part of the CMD can be interpreted as 
tracers of a very young population located in a spiral arm.
 The idea of a spiral
arm is strongly supported by the fact that we find this young population in 
the fields F3 and F4, which are in the direction of the Sgr-Car arm
(see Fig. \ref{find} and \ref{aufbau}), whereas the fields F1 and F2, which
are not expected to cross the spiral arm,
 do not show any evidence for such a population.

The spiral arm has been parameterized following 
\cite{Vall95} where however the parameters
of the spiral pattern are set to reproduce our data.
  We adopt an arm/inter-arm density ratio of 2
as found in external galaxies  (Rix \& Zaritsky 1995) and suggested
for our own Galaxy by Efremov (1997). 
The spiral arm is supposed to have a Gaussian distribution with a
$\sigma$ of 300 pc as suggested by \cite{TayCor93}, and \cite{Vall95}.
We impose that the age of the spiral arm population is younger than
1 $\times 10^8 $ y. 
%%  On the basis of dynamical considerations, some $10^8$ yr
%% is the time requested for a star born inside the arm with to leave the arm.
From CMD simulations, we find out that the
 faintest magnitude of the blue population
gives hints about the distance of the spiral arm from us
(see Fig \ref{CMD_simulation}).
 We find out that the distance of the spiral arm
in the direction of F3 is about 1.3 $\pm 0.1$\,kpc, with the maximum of the
distribution at 2\,kpc. 
In the direction of F4, we find that
the spiral arm is at a distance of 1.5 $\pm 0.2$\,kpc. 
These results are of course dependent on the adopted
parameterization of the spiral arm. They are 
consistent with the spiral arm pattern
defined by the pulsar distribution (\cite{TayCor93})
or derived from optical observations for the local solar environment
(\cite{Hum76}).

\section {Conclusions}
\label{sec_conclusion}

Deep CCD photometry in the V and I pass-bands is presented
for 4 low latitude Galactic fields.
The studied fields are located between l=265 and l=305.

The scale height and the scale lenght of the various populations
are derived. The data seem to favor a disk scale height around 
$h_z = 250\pm 0.60\rm\,pc$ and a scale lenght $>$ 1100 pc.
The star formation seems to have been decreasing in time,
while  a constant or increasing star formation rate as the one
found
using Hipparcos data in the solar neighborhood, cannot reproduce
the features of the CMDs: too many blue stars are expected.

From the star-counts in the studied  fields
we derive the local mass density in stars more massive than 0.1 M$_\odot$,
 found to be in the range
0.036-0.022 M$_\odot$ pc$^-3$.

In two fields, namely F3 and F4 located at l=292 and l=305
evidence is found of the presence of a very
 young population. The data can be reproduced
imposing the presence of a spiral arm. In these direction the
Sgr-Car arm is crossed at a distance of 1.3$\pm 0.1$ and 1.5 $\pm 0.2$\,Kpc
respectively. These results are consistent with 
previously defined spiral arm patterns
(\cite{TayCor93}; \cite{Hum76}).

Due to the small field of view, two of the studied fields, F1 and F2
do not set strong constrains on the scale height and the scale lenght
of the disk. 
 A larger field of view (see e.g.
the WFI at the 2.2m ESO telescope having 30$^\prime \times$ 30$^\prime$)
would allow us to have good statistics down to faint magnitudes.
At l=305, b=-4.8  more than 10$^4$ disk stars
down to V=21 are expected in such a large field. 
In a forthcoming paper we will discuss the Galactic structure
towards the center making use of the good statistics of the WFI data.

\acknowledgement{
%%%% A.V. thanks  the Graduirten
%%% Kollege  and the Sternwarte (Bonn) for the kind ospitality
%%% in the period while the paper was written
The paper was written while A.V. was visiting scientist of the
Graduierten Kollege at the Sternwarte der Universit\"at (Bonn). L.S.
was supported by the Deutsche Forschungsgemeinschaft DFG, under the grant
N$^0$ Schm 1444/1-1.}

\end{document}